\documentclass{ws-rv975x65}
\usepackage{subfigure}     
\usepackage{ws-rv-van}     
\usepackage{psfrag}
\usepackage{colortbl}

\usepackage[spanish]{babel}
\makeindex

\newcommand\T{\rule{0pt}{2.6ex}}
\newcommand\B{\rule[-1.7ex]{0pt}{0pt}}

\numberwithin{equation}{section}

\def\ut#1{\rlap{\lower1ex\hbox{$\sim$}}#1{}}
\newcommand{\N}{\mathbb{N}}
\newcommand{\C}{\mathbb{C}}
\newcommand{\R}{\mathbb{R}}
\newcommand{\be}{\nopagebreak[3]\begin{equation}}
\newcommand{\ee}{\end{equation}}
\newcommand{\ba}{\nopagebreak[3]\begin{eqnarray}}
\newcommand{\ea}{\end{eqnarray}}
\DeclareFontFamily{U}{rsfs}{}         
\DeclareFontShape{U}{rsfs}{m}{n}{<5> rsfs5 <6><7> rsfs7          %
  <8><9><10><10.95><12><14.4><17.28><20.74><24.88> rsfs10}{}     %
\DeclareMathAlphabet{\mathfs}{U}{rsfs}{m}{n}                     %
\newcommand{\mfs}[1]{\mathfs {#1}}                               %
\newcommand{\n}{{\nonumber}}
\newcommand{\va}{\scriptscriptstyle}

\newcommand{\sL}{{\mfs L}}

\newcommand{\sZ}{{\mfs Z}}

\newcommand{\sI}{{\mfs I}}\newcommand{\sO}{{\mfs O}}

\newcommand{\nn}{\sqrt{j(j+1)}}

\def\pb#1{\rlap{\lower1.5ex\hbox{$\longleftarrow$}}{#1}}
\def\dpb#1{\rlap{\lower1.5ex\hbox{$\Longleftarrow$}}{#1}}
\def\spb#1{\rlap{\lower1.5ex\hbox{$\leftarrow$}}{#1}}
\def\sdpb#1{\rlap{\lower1.5ex\hbox{$\Leftarrow$}}{#1}}

\begin{document}

\chapter[Quantum Geometry and Black Holes]{Quantum Geometry and Black Holes}\label{ra_ch1}

\author[J. Fernando Barbero G. and Alejandro Perez]{J. Fernando Barbero G.}

\address{Instituto de Estructura de la Materia, CSIC, Serrano 123, 28006 Madrid, Spain.\footnote{Grupo de Teorias de Campos y Fisica Estadistica, Instituto Universitario Gregorio Mill\'an Barbany, Universidad Carlos III de Madrid, Unidad Asociada al IEM-CSIC.}}

\author[J. Fernando Barbero G. and Alejandro Perez]{Alejandro Perez}

\address{Centre de Physique Th\'eorique, Campus de Luminy, 13288 Marseille, France.\footnote{Unit\'e Mixte de Recherche (UMR 6207) du CNRS et des Universit\'es Aix-Marseille I, Aix-Marseille II, et du Sud Toulon-Var; laboratoire afili\'e \`a la FRUMAM (FR 2291).}}

\begin{abstract}
We present an overall picture of the advances in the description of black hole physics from the perspective of loop quantum gravity. After an introduction that discusses the main conceptual issues we present some details about the classical and  quantum geometry of isolated horizons and their quantum geometry and then use this scheme to give a natural definition of the entropy of black holes. The entropy computations can be neatly expressed in the form of combinatorial problems solvable with the help of methods based on number theory and the use of generating functions. The recovery of the Bekenstein-Hawking law and corrections to it is explained in some detail. After this, due attention is paid to the discussion of semiclassical issues. An important point in this respect is the proper interpretation of the horizon area as the energy that should appear in the statistical-mechanical treatment of the black hole model presented here. The chapter ends with a comparison between the microscopic and semiclassical approaches to the computation of the entropy and discusses a number of issues regarding the relation between entanglement and statistical entropy and the possibility of comparing the subdominant (logarithmic) corrections to the entropy obtained with the help of the Euclidean path integral with the ones obtained in the present framework.

\end{abstract}

\body

\clearpage

{\tiny \noindent {\em Detr\'as de cada espejo\\
hay una estrella muerta\\
y un arco iris ni\~no\\
que duerme.\\
\vskip.2cm

\noindent Detr\'as de cada espejo\\
hay una calma eterna\\
y un nido de silencios\\
que no han volado.\\

\ \ \ \  \ \ \ \  F.G. Lorca}}
\section{Discussion of the conceptual issues}\label{ra_sec1}

Black holes are remarkable solutions of classical general relativity describing important aspects of the physics
of gravitational collapse. Their existence in our nearby
universe is  supported by a great amount of observational evidence \cite{Narayan:2013gca}. When isolated, these systems
are expected to be simple for late and distant observers.
Once the initial very dynamical phase of collapse has passed, the system should settle
down to a stationary situation completely described by
the Kerr-Newman solution\footnote{Such scenario is based on physical grounds, some concrete indications from 
perturbation theory, and the validity of the  so called no-hair theorem\cite{}.} labelled by three macroscopic parameters: 
the mass $M$, the angular momentum $J$, and the electromagnetic charge $Q$.

The fact that the final state of gravitational collapse is described by only three macroscopic parameters, independently of the
details of the initial conditions leading to the collapse, could be taken as a first indication of the thermodynamical nature of black holes (which as we will see below is really of quantum origin).
In fact the statement in the first paragraph contains the usual coarse graining perspective of thermodynamical physics in the
assertion that  for sufficiently long  times after collapse {\em the system should settle
down to a stationary situation... described by three parameters}. The details about how this settling down takes place depend indeed on the initial conditions leading to 
the collapse (the microstates of the system). The coarse graining consists of neglecting these details in favour of the idealisation of stationarity.  

Another classical indication is Hawking area theorem\cite{Hawking:1971tu} stating that for mild energy conditions (satisfied by classical matter fields)
the area of a black hole horizon can only increase in any physical process. Namely, the so-called second law of black hole mechanics holds:
\be\label{2nd}
\delta A\ge 0.
\ee
This brings in the irreversibility characteristic of thermodynamical systems to the context of black hole physics and motivated Bekenstein\cite{Bekenstein:1973ur}
to associate with BHs a notion of entropy proportional so their area.
Classically, one can also prove the so-called first law of BH mechanics\cite{Bardeen:1973gs}
relating different nearby stationary BH spacetimes of
Einstein-Maxwell theory
\begin{eqnarray}\label{1st}
\delta M=\frac{\kappa}{8\pi} \delta A +\Omega \delta J+\Phi  \delta Q,
\end{eqnarray}
where $\Omega$ is the angular velocity of the horizon, $\Phi$ is horizon electric potential, and $\kappa$ is the surface gravity. 

%
The realization that black holes can indeed be considered (in the semiclassical regime) as thermodynamical systems came with the
discovery of black hole radiation\cite{Hawking:1974sw}. In the mid 70's Hawking considered the scattering of a quantum test field on a space time background geometry representing gravitational collapse of a compact source. Assuming that very early observers far away from the source prepare the field in the vacuum state he showed that, after the very dynamical phase of collapse is replaced by a stationary quasi equilibrium situation, late observers in the future measure an afterglow of particles of the test field coming from the horizon with a temperature
\be
T_{H}=\frac{\kappa}{2\pi}.
\ee  
As black holes radiate, the immediate conclusion is that they must evaporate through the (quantum phenomenon of) emission of Hawking radiation. The calculation of Hawking neglects such back reaction
but provides a good approximation for the description of black holes that are sufficiently large, for which the radiated power is small relative to the scale defined by the mass of the black hole. These black holes are referred to as {\em semiclassical} in this chapter.

This result, together with the validity of the first and second laws, suggest that semiclassical black holes should have an  associated  
entropy (here referred to as the Bekenstein-Hawking entropy) given by 
\be\label{entro0}
S_{H}=\frac{A}{4\ell_{Pl}^2}+S_0
\ee
where $S_0$ is an integration constant that cannot be fixed by the sole use of the first law. 
In fact, as in any thermodynamical system, entropy cannot be determined only by the use of the first law. 
Entropy can either be measured in an experimental setup (this was the initial way in which the concept was introduced) or calculated from 
the basic degrees of freedom by using statistical mechanical methods once a model for the fundamental building 
blocks of the system is available.

More precisely, even though the thermodynamical nature of semiclassical black holes is a
robust prediction of the combination of general relativity and quantum field theory as a first approximation to quantum gravity, 
the precise expression for the entropy of black holes is a question that can only be answered within the
framework of quantum gravity in its semiclassical regime. This is a central question for any 
proposal of quantum gravity theory. 

%
This chapter will mainly deal with the issue of computing 
black hole entropy for semiclassical black holes which, as we will argue here, already presents an important challenge
to quantum gravity but seems realistically within reach at the present stage of development of the approach. 
The formalism applies to physical black holes of the kind that can be formed in the early primordial universe or other 
astrophysical situations (no assumption of extremality or supersymmetry is needed).

Questions related to the information loss paradox, or the the fate of unitarity are all issues 
that  necessitate full control of the quantum dynamics in regimes far away from the semiclassical one. For that reason we designate this set of questions as the {\em hard problem}.
These involve in particular the understanding of the dynamics near and across  (what one would classically identify with) the interior singularity. There are studies of the quantum dynamics through models near the (classically apparent) singularities
of general relativity indicating that not only the quantum geometry is well defined at the classically pathological regions, but also the
quantum dynamics is perfectly determined across them. For the variety of results concerning cosmological singularities we refer the reader to {\bf Chapter LOOP QUANTUM COSMOLOGY}.
Similar results have been found in the context of black holes \cite{Ashtekar:2005qt}. These works indicate that singularities are generically avoided due to quantum effects
at the deep Planckian regime.  Based on these results new paradigms have been put forward  concerning the {\em hard problem}\cite{Ashtekar:2005cj}.
 The key point is that the possibility of having physical dynamics beyond the apparent classical singularities 
 allows for information to be lost into causally disconnected worlds (classical singularities as sinks of information) or to be recovered in subtle ways during and after evaporation as suggested by results in 2d black hole systems\cite{Ashtekar:2008jd, Ashtekar:2010hx, Ashtekar:2010qz}. All these scenarios would be compatible with a local notion of unitarity\cite{Wald:1999vt}. The information paradox could also be solved \cite{Perez:2014ura, Perez:2014xca} if quantum correlations with the (discrete) UV Planckian degrees of freedom remain hidden to low energy (semiclassical) observers. This possibility is appealing in an approach such as LQG where continuum space-time is obtained by coarse graining \cite{Varadarajan:1999it, Ashtekar:2001xp, Ashtekar:2002sn}. Space limitations prevent us from discussing the {\em hard problem} further in this chapter.


To date, investigations within the LQG framework,  can be divided into the following categories:
isolated horizons and their quantum geometry (Sections \ref{IH} and \ref{IH_quantum geometry}); rigorous counting of micro-states
(Section \ref{IH_counting}); semiclassical quasi-local formulation  (Section \ref{semi}); spin foam dynamical accounts and low energy dynamical counterparts (Section \ref{sf}); and the
Hawking effect phenomenology and insights from symmetry reduced models (Section \ref{HR}).
The different sections are largely self-contained  so they can be read independently.



\section{Isolated horizons}\label{IH}

The model employed to describe black holes in loop quantum gravity is based on the use of \textit{isolated horizons} (IH), a concept introduced around the year 2000 by Ashtekar and collaborators \cite{ABF,AFK0,ABL1,AFK} and developed by a number of other researchers \cite{Lew1,LP1}.\footnote{The mathematical foundations of the subject were developed by Kupeli in \cite{Kupeli}.} The main goal of this line of work was to find a \textit{quasilocal} notion of horizon that could be used in contexts were the teleological nature of event horizons (i.e. the need to know the whole spacetime in order to determine if they are present) is problematic.

The most important features of isolated horizons are: their quasilocal nature, the availability of a Hamiltonian formulation for the sector of general relativity containing IH's, the possibility of having physically reasonable versions of some of the laws of black hole thermodynamics and the existence of quasilocal definitions for the energy and angular momentum. It is important to remark, already at this point, the striking interplay between the second and the third issues.

The quasilocality of isolated horizons reflects itself in the fact that they can be described by introducing an inner spacetime boundary and imposing boundary conditions on the gravitational field defined on it (either in a metric or a connection formulation). As we want to describe black holes in equilibrium, it is natural to look for particular boundary conditions compatible with a static horizon but allowing the geometry outside to be dynamical (admitting, for example, gravitational radiation). This will lead us to consider a sector of general relativity significantly larger than the one consisting of standard black hole solutions.

The sector of the gravitational phase space that we will be dealing with admits a Hamiltonian, hence, it is conceivable to quantize it to gain an understanding of quantum black holes. This is one of the advantages of working with isolated horizons and a very non-trivial fact because such a Hamiltonian formalism is not available for other sectors of general relativity.  The approach that we will follow is somehow reminiscent of the study of symmetry reductions of general relativity (mini and midisuperspaces). As the sector of the phase space of the reduced system is large enough---actually infinite dimensional---it seems reasonable to expect that the quantum model that we consider will provide a good physical approximation for the equilibrium phenomena that we want to discuss\footnote{The quasilocal description of \textit{dynamical} black hole behaviors can be achieved by using the so called dynamical horizons \cite{AK2}.}, in particular the microscopic description of black hole entropy and the Bekenstein-Hawking area law.

We review next the construction of isolated horizons justifying, along the way, the conditions that have to be incorporated during the process. The main results regarding the geometry of isolated horizons can be found in \cite{AKLR}. We will be defining different types of null hypersurfaces until we arrive at the concept of isolated horizon. In the process we will introduce the notation that will be used in the following.

\textit{Null hypersurfaces:} Let $\mathcal{M}$ be a 4-dim manifold and $g_{\mu\nu}$ a Lorentzian metric on $\mathcal{M}$. A 3-dimensional embedded submanifold $\Delta\subset\mathcal{M}$ will be called a null hypersurface if the pull-back $g^{\Delta}_{ab}$ of $g_{\mu\nu}$ onto $\Delta$ is degenerate. This condition implies the existence of a null normal $\ell^a$ \textit{tangent} to $\Delta$. Notice that there is not a \textit{unique} projection of tangent vectors $X^\mu$ sitting on $p\in\Delta$ onto the tangent space $T_p\Delta$ and, hence, it is impossible to define an induced connection on $\Delta$.

\textit{Non-expanding null hypersurfaces:} The degeneracy of the metric $g^{\Delta}_{ab}$ implies that there is not a \textit{unique} inverse metric, however it is always possible to find $g_\Delta^{ab}$ such that $g_{ab}^\Delta=g_{aa'}^\Delta g_\Delta^{a'b'}g_{b'b}^\Delta$. If $\ell^a$ is a field tangent to $\Delta$ consisting of null normals, we define its \textit{expansion} $\theta_\ell$ associated with a particular choice of $g_\Delta^{ab}$ as $\theta_\ell:=g_\Delta^{ab}\mathcal{L}_\ell g^\Delta_{ab}$. The invariance under rescalings implies that this expansion cannot be associated in an intrinsic way to the null hypersurface $\Delta$ unless it is zero. Null hypersurfaces with zero expansion in the previous sense will be referred to as \textit{non-expanding}.

\textit{Non-expanding horizons (NEH):} As we want to model black holes in four dimensions --for which the horizons have a simple geometry-- we will require that: i) $\Delta$ is diffeomorphic to $\mathbb{S}^2\times (0,1)$ where $\mathbb{S}^2$ is a 2-sphere. ii) For each $x\in \mathbb{S}^2$ this diffeo maps $\{x\}\times(0,1)$ to null geodesics on $\Delta$. iii) For each $t\in(0,1)$, $\mathbb{S}^2\times \{t\}$ is mapped onto a spacelike 2-surface in $\Delta$. We impose now a key physical condition by requiring that the metric $g_{\mu\nu}$ be a solution to the Einstein field equations and demanding that the pull back of the stress-energy-momentum tensor $T_{\mu\nu}$ on $\Delta$ satisfies the condition $T^\Delta_{ab}\ell^a\ell^b\geq0$. This is equivalent to the condition $R^\Delta_{ab}\ell^a\ell^b\geq0$ on the pull-back of the Ricci tensor to $\Delta$. The preceding conditions on non-expanding null surfaces define \textit{non-expanding horizons}. An important feature of them is that, as a consequence of the non-expansion condition,  cross sections are marginally trapped surfaces and have constant area. Also, the Raychaudhuri equation together with the non-expanding condition implies that $R^\Delta_{ab}\ell^a\ell^b=T^\Delta_{ab}\ell^a\ell^b=0$ and $\mathcal{L}_\ell g^\Delta_{ab}=0$. This can be interpreted as the fact that non-expanding horizons \textit{are in equilibrium}.

\textit{Weakly isolated horizons (WIH):} In order to incorporate the laws of black hole mechanics to the present quasilocal framework we need to add additional structure to the preceding constructions. For example, the notion of temperature for ordinary black holes relies on the concept of surface gravity $\kappa$ (see, for example, \cite{Wald}). We can introduce now a rather similar concept by imposing additional requirements to NEH's. Given a non-expanding horizon, it can be shown [see \cite{Kupeli,AKLR,AFK,ABL1}] that the spacetime connection $\nabla$ induces a unique connection $\mathcal{D}$ compatible with the induced metric $g^\Delta_{ab}$. We also declare as equivalent all the normal null fields related by constant rescalings (we denote these equivalence classes as $[\ell]$). A weakly isolated horizon is now a pair $(\Delta,[\ell])$ consisting of a non-expanding horizon $\Delta$ and class of null normals $[\ell]$ such that
\begin{equation}
(\mathcal{L}_\ell \mathcal{D}_a-\mathcal{D}_a\mathcal{L}_\ell)\ell^b=0\,.
\label{IH_001}
\end{equation}
Geometrically this requirement is equivalent to the condition that \textit{some} components of the connection defined by $\mathcal{D}$ are left invariant by the diffeos defined by $\ell$ on $\Delta$ (or roughly speaking are ``time independent''). This means that  $\nabla_{\ell}\ell=\kappa \ell$ with $\kappa$ \textit{constant} for each $\ell\in[\ell]$. It is important to mention here that different choices of $[\ell]$ lead to inequivalent weakly isolated structures on the same non-expanding horizon.

\textit{Isolated horizons (IH):} Isolated horizons are weakly isolated horizons $(\Delta,[\ell])$ for which
\begin{equation}
(\mathcal{L}_\ell \mathcal{D}_a-\mathcal{D}_a\mathcal{L}_\ell)\tau^b=0\,,
\label{IH_002}
\end{equation}
for every tangent field $\tau^a$ on $\Delta$. This condition can be read as $[\mathcal{L},\mathcal{D}]|_\Delta=0$. An important difference between WIH's and IH's is that, whereas a given non expanding horizon admits infinitely many weakly isolated horizon structures, for an isolated horizon the only freedom in the choice of null normals consists of constant rescalings \cite{AKLR}.

Non-trivial examples of all these types of horizons can be found in the extensive literature available on the subject (see \cite{AKLR} and references therein); in any case it is important to keep in mind that any Killing horizon diffeomorphic to $\mathbb{S}^2\times\mathbb{R}$ is an isolated horizon so the concept is a genuine --and useful-- generalization that encompasses all the globally stationary black holes.

Multipole moments can be used to define spherically symmetric isolated horizons in an intrinsic way\cite{Ashtekar:2004gp, Ashtekar:2004nd}. They are useful because, for stationary spacetimes in vacuum, they determine the near horizon geometry. Concrete expressions for these objects can be written in terms of the  $\Psi_2$ Newmann-Penrose component of the Weyl tensor. In the spherically symmetric case $\mathrm{Im}\Psi_2=0$ and $\mathrm{Re}\Psi_2$ is constant which implies that the only non-zero multipole moment is $M_0$. This condition provides the intrinsic characterization mentioned above.

The zeroth and first laws of black hole mechanics have interesting generalizations for weakly isolated and isolated horizons. In the case of the zero law the geometric features of weakly isolated horizons guarantee that a suitable concept of surface gravity can be introduced. This is done as follows \cite{AKLR}. For a non expanding horizon $\Delta$ the null normal $\ell^a$ has vanishing expansion, shear and twist. It is then straigthforward to show that there must exist a 1-form $\omega_a$ on $\Delta$ such that $\nabla_a\ell^b=\omega_a\ell^b$ and $(\mathcal{L}_\ell\omega)_a=0$ (the last condition as a consequence of the definition of weakly isolated horizon). Defining now the surface gravity associated with the null normal $\ell^a$ as $\kappa_\ell:=\ell^a\omega_a$ we have $\displaystyle d\kappa_\ell=d(\omega_a\ell^a)=(\mathcal{L}_\ell\omega)_a=0$ and hence $\kappa_\ell$ is constant on the horizon. This is analogous to the behavior of the surface gravity for Killing horizons and provides us with the sought for generalized zeroth law.

The generalization of first law of black hole dynamics requires the definition of a suitable energy associated with the isolated horizon. A way to proceed is to look for a Hamiltonian description for the sector of general relativity containing IHs. The availability of such a formulation is a very non-trivial and remarkable fact, and it is a necessary first step towards quantization. In generally covariant theories the Hamiltonian generating time translations is given by a surface integral (once the constraints are taken into account). In the present case there will be, hence, an energy associated with the isolated horizon (and an extra ADM term corresponding to the boundary at infinity).\footnote{Similar arguments apply to the angular momentum} In practice, associating a Hamiltonian to the boundary $\Delta$ requires the choice of an appropriate concept of time evolution defined by vector field $t^a$ with appropriate values $t^a_\Delta$ at the horizon. In simple examples (for instance, non rotating isolated horizons) it is natural to take $t^a_\Delta$ proportional to the null normal $\ell^a$, however, there is some freedom left in the choice of $\ell^a$ by the IH boundary conditions. By choosing $t^a_\Delta$ in such a way that the surface gravity is a specific function of the area (and other charges) and demanding that the evolution generated by $t^a_\Delta$ is Hamiltonian\cite{AKLR} one gets the first law as a necessary and sufficient condition. In this way, there is a family of mathematically consistent first laws parametrised by these choices.

The textbook approach to obtain the Hamiltonian would consist in starting from a suitable action principle for general relativity in a spacetime manifold with an inner boundary where the isolated horizon boundary conditions are enforced. This action can be written in principle both in terms of connection or metric variables. The standard Dirac approach to deal with constrained systems (or more sophisticated formalisms such as the one given in \cite{GNH}) can then be used to get the phase space of the model, the symplectic structure, the constraints and the Hamiltonian \cite{Booth}. Notice that owing to the presence of boundaries one should expect, in principle, a non zero Hamiltonian consisting both in horizon contributions (defining the horizon energy $E_\Delta$ in terms of which the first law is spelled) and the standard ADM energy associated with the boundary at infinity. A different approach that has some computational advantages relies on the covariant methods proposed and developed in \cite{Witten,Crnk}. Their essence is to directly work in the space of solutions to the Einstein field equations with fields subject to the appropriate boundary conditions (in particular the isolated horizon ones). Despite the fact that the solutions to the field equations in most field theories are not known it is possible to obtain useful information about the space of solutions and, in particular, the symplectic form defined in it.

As this is a crucial ingredient to understand the quantization of the model and the quantum geometry of isolated horizons we sketch now the derivation of the symplectic structure based on covariant phase space methods. Let us suppose that we have a local coframe $e_\mu^I\,,\quad I=1,\ldots,4$ in the spacetime\footnote{$\eta_{IJ}$ denotes the Minkowski metric. In the following we will suppress spacetime indices when working with differential forms} $(\mathcal{M},g_{\mu\nu})$ and the frame connection $\Gamma^I_{J}$ defined by $d e^I+\Gamma^I_{J}\wedge e_{\nu}^J=0$ with $\Gamma_{IJ}+\Gamma_{JI}=0$. If we denote tangent vectors (at a certain solution $e^I$) as $\delta e^I$ it is straightforward to show that the 3-form defined on $\mathcal{M}$ by
\begin{equation}
\omega(\delta_1,\delta_2):=\frac{1}{2}\varepsilon_{IJKL}\delta_{[1}(e^I\wedge e^J)\wedge \delta_{2]}\Gamma^{KL}-\frac{1}{\gamma}\delta_{[1}(e^I\wedge e^J)\wedge \delta_{2]}\Gamma_{IJ}
\label{IH_003}
\end{equation}
is closed if $e^I$ is a solution to the Einstein field equations (in the previous expression $\gamma$ is the Immirzi parameter). This means that if we have two 3-surfaces $\Sigma_1$ and $\Sigma_2$ defining the boundary of a 4-dim submanifold of $\mathcal{M}$ then
\begin{equation}
\Omega(\delta_1,\delta_2)=\int_{\Sigma}\omega(\delta_1,\delta_2)
\label{symp_1}
\end{equation}
is independent of $\Sigma$. If an inner boundary, such as an isolated horizon, is present then a similar argument leads to the obtention of the symplectic form. Indeed, let us take a region of $\mathcal{M}$ with an inner boundary $\Delta$ (a causal 3-surface) and a family of spatial 3-surfaces $\Sigma$ such that every pair $\Sigma_1$ and $\Sigma_2$, defines a 4-dim spacetime region bounded by $\Sigma_1$, $\Sigma_2$ and the segment of the surface $\Delta$ contained between the 2-surfaces $\Sigma_1\cap \Delta$ and $\Sigma_2\cap \Delta$. Let us suppose also that, for every pair of tangent vectors $\delta_1,\delta_2$, there is a 2-form $\alpha(\delta_1,\delta_2)$ on $\Delta$ such that the pullback of $\omega$ onto $\Delta$ is exact [$\omega^\Delta(\delta_1,\delta_2)=d\alpha(\delta_1,\delta_2)$]. When these conditions are satisfied it is possible to generalize (\ref{symp_1}) in such a way that, in addition to the bulk term obtained above, it also has a surface term and the resulting expression is still independent of the choice of $\Sigma$
\begin{equation}
\Omega(\delta_1,\delta_2)=\int_{\Sigma}\omega(\delta_1,\delta_2)+\int_{\Sigma\cap\Delta}\alpha(\delta_1,\delta_2)\,.
\label{symp_2}
\end{equation}
These types of surface terms are defined both for weakly isolated horizons or spherical isolated horizons as inner boundaries. In the case of weakly isolated horizons of fixed area $A$ it is possible to perform a gauge fixing such that the only symmetry left is a $U(1)$ symmetry. In such a situation it is possible to see that the surface contribution to (\ref{symp_2}) has the form

\begin{equation}
\frac{A}{\pi\gamma}\int_S \delta_{[1} V\wedge \delta_{2]}V
\label{IH_004}
\end{equation}
where $V$ is a $U(1)$ connection on the spheres $S$ that foliate the horizon. It is important to notice that this is a $U(1)$ Chern-Simons symplectic form. It is convenient now to rewrite the bulk term by using Ashtekar variables as
\begin{equation}
2\int_\Sigma \delta_{[1}E^a_i\wedge\delta_{2]}A^i_a\,.
\label{IH_005}
\end{equation}
It is necessary to mention at this point\cite{Ashtekar:1999wa} that the values of the $U(1)$ connection and the pullbacks of the connection/triad variables are not independent but are connected through a horizon constraint of the form\footnote{Here $r^i$ denotes a fixed internal vector and we have used units such that $8\pi G=1$.}
\begin{equation}\label{U1Conds}
\left.(dV)_{ab}+\frac{2\pi\gamma}{A}\epsilon_{abc}(E^c_i r^i)\right|_{\Delta}=0\,.
\end{equation}
The quantum version of this condition plays a central role in the quantization of this model.

For spherical isolated horizons it is possible to define the Hamiltonian framework without gauge fixing on the horizon \cite{Engle:2009vc,Engle:2010kt,Engle:2011vf}. In such formulation the symplectic form in the field space has an $SU(2)$ Chern-Simons surface term of the form
\begin{equation}
\frac{A}{8\pi^2(1-\gamma^2)\gamma}\int_S \delta_1 A_i\wedge \delta_2 A_i\,,
\label{IH_006}
\end{equation}
where $A_i$ denotes the pullback of the $SU(2)$ connection to the horizon. Now the horizon constraint is not a single condition but the three conditions
\begin{equation}\label{SU2Conds}
\left.\frac{1}{2}\epsilon_{abc}E^{ci}+\frac{A}{8\pi^2(1-\gamma^2)\gamma}F_{ab}^i\right|_{\Delta}=0\,,
\end{equation}
written in terms of the curvature $F_{ab}^i$. The difference between the $U(1)$ and the $SU(2)$ approaches stems, mainly, from this fact but it is important to mention that the physical assumptions used to define both models are slightly different. 

A remark is in order when comparing equations (\ref{U1Conds}) and (\ref{SU2Conds}). At first sight there seems to be a mismatch in the number of conditions. In fact, as a consequence of the gauge fixing that reduces the  $SU(2)$ triad rotation in the bulk to $U(1)$ on the boundary, one has two extra conditions on the fluxes corresponding to \be\label{orthog} E^c_i y^i=0=E^c_i x^i\ee where $x^i$ and $y^i$ are internal directions orthogonal to each other and to $r^i$. We see that the three conditions in (\ref{SU2Conds}) are recovered. Due to the non commutativity of the $E^c_i$ the previous conditions cannot be satisfied in the quantum theory: only (\ref{U1Conds}) is imposed in the $U(1)$ framework. As a result the $U(1)$ framework slightly over counts  states,  a fact which (under qualifications that are discussed at the end of Section \ref{IH_quantum geometry}) is reflected in the form of logarithmic corrections to the micro canonical entropy (see table in Section \ref{IH_counting}).


\section{Quantum geometry of Weakly Isolated Horizons}\label{IH_quantum geometry}

The formulation put forward in the preceding section can be used to identify the degrees of freedom that account for the black hole entropy and understand their quantum origin. It is precisely the quantum geometry associated with weakly isolated horizons { that} will let us understand the origin of black hole entropy in the LQG framework. As we will discuss in this section a special role will be played by the quantum horizon boundary conditions. For simplicity of exposition we will restrict ourselves to the setting provided by Type I WIH's and {suppose} that we do not have matter nor extra charges. The starting point of the following construction is a WIH of fixed area\footnote{And fixed charges, in general.} $a$. As we mentioned in the preceding section the sector of general relativity consisting of solutions to the Einstein field equations on regions bounded by weakly isolated horizons admits a Hamiltonian formulation so that its quantization can be considered in principle. It is important to point out, however, that the following construction must be based on the use of connection-triad variables of the Ashtekar type ({\bf see Chapter LOOP QUANTUM GRAVITY}). To our knowledge such a construction is not available in the geometrodynamical framework \footnote{As mentioned in Section \ref{IH} there is a boundary contribution to the gravity symplectic form that can be written as an $SU(2)$ Chern-Simons symplectic form in connection variables. In triad variables $e_a^i$ this is\cite{Engle:2010kt}
$\gamma^{-1}\kappa^{-1}\int \delta_1 e_i\wedge\delta_2 e^i $. If we define smeared fluxes in the usual way $E(S,\alpha)=\int_S \epsilon_{ink}\alpha^i e^j\wedge e^k$ then it follows that $
\{E(S,\alpha),E(S',\beta)\}=\gamma \kappa E(S\cap S',[\alpha,\beta])$.
The previous non commutativity of fluxes is characteristic of  the bulk holonomy flux algebra\cite{Ashtekar:1998ak}. 
}. One of the reasons for this is the central role that Chern-Simons theories play in the following arguments.

We have chosen to describe with some degree of detail  the $U(1)$ 
gauge fixed formulation of quantum IHs. The $SU(2)$ invariant framework\cite{Engle:2009vc, Engle:2010kt} can be constructed along similar lines.  In addition to the point discussed at the end of the previous section, the main advantage of the latter is that both boundary and bulk fields possess the same gauge symmetry. This allows the IH quantum constraints to be interpreted as first class constraints generating the common symmetry\cite{Engle:2010kt}. Results of the $SU(2)$ invariant formulation will be presented without details at end of the section.  

A very striking feature of the construction that we have discussed at the end of the preceding section is the presence of a surface term in the symplectic structure. Such surface terms are usually absent for field theories with boundaries (at least in simple models, see \cite{nos}). They are a very distinctive feature of the present approach. From the point of view of the quantization of the model this surface term strongly suggests the necessity to introduce a Hilbert space associated with the boundary. The fact that it corresponds to a Chern-Simons model directly leads to the consideration of a Chern-Simons quantization.

In statistical mechanics, the classical and quantum degrees of freedom that account for the entropy of a thermodynamical system are usually the same. For example the atoms in a gas, interpreted as point particles in a box, are in one to one correspondence with the quantum degrees of freedom used to model the gas as an ensemble of particles in an infinite potential well. In the present case the logical interpretation of the results about the specification of spacetimes with isolated horizons \cite{Lew1} implies that there are no classical degrees of freedom associated with them. What is then the origin of the entropy of black holes in this setting? The answer lies in the nature of equations (\ref{U1Conds}) and (\ref{SU2Conds}). More precisely, the intersections of the edges of the spin network (excitations of the field $E^a_i$) used to represent a suitable quantum bulk state are treated as point particle defects at the horizon---effectively excising them. The degrees of freedom  of the horizon Chern-Simons theory created in this way are  responsible for the entropy.

The construction of the LQG Hilbert spaces has been reported in {\bf Chapter LOOP QUANTUM  GRAVITY}. In the present context we will import results from these constructions --for the bulk degrees of freedom-- and also from the quantization of Chern-Simons theories to deal with the { horizon\cite{ABK}}. As mentioned before it is natural to introduce a Hilbert space $\mathcal{H}=\mathcal{H}_S\otimes\mathcal{H}_V$ where the Hilbert spaces $\mathcal{H}_{Hor}$ and $\mathcal{H}_{Bulk}$ are associated with the horizon and the bulk spacetime respectively.

The volume or bulk Hilbert space $\mathcal{H}_{Bulk}$
is a subspace of the usual LQG Hilbert space $L^2(\bar{\mathcal{A}},\mu_{AL})$ defined in a suitable space of generalized connections with the help of the uniquely defined Ashtekar-Lewandowski measure { (see {\bf Chapter LOOP QUANTUM  GRAVITY})}. A useful orthonormal basis for this type of Hilbert space is provided by \textit{spin networks} with edges that (may) { transversally} pierce the inner spacetime boundary that models the black hole.  These points will be referred to as { \textit{punctures}}; they are endowed with the quantum numbers that label the edges defining them. By using these punctures it is possible to represent the bulk Hilbert space as an { orthogonal} sum \cite{Domagala:2004jt}
\begin{equation}
\mathcal{H}_{Bulk}=\bigoplus_{(P,j,m)}\mathcal{H}^{P,j,m}_{Bulk}
\label{QG_001}
\end{equation}
{ extended} to all the possible finite sets $P=\{P_1,\ldots,P_n\}$ consisting of points at the spherical sections of the { horizon. The} $(j,m)$ labels correspond to edges piercing the horizon transversally and the empty set corresponds to spin networks that do not pierce the horizon.

In order to construct the surface Hilbert space it is necessary to excise the punctures from the sphere $S$ at the horizon and study the quantization of a Chern-Simons model in the { resulting punctured surface}. From a classical point of view this modification of the horizon topology has the effect of introducing topological degrees of freedom in the model { (that can be thought of as the holonomies around closed loops surrounding the punctures of the, otherwise, flat connection)}, however one has to keep in mind that these punctures are induced by spin network states defined in the bulk, hence, they have a { quantum} origin.

The Chern-Simons quantization requires us to impose a prequantization condition on the classical horizon area. In the present situation { it} reads \cite{ABK} $A_\kappa=4\pi\gamma \ell_{Pl}^2 \kappa$ with $\kappa\in\mathbb{N}$. In analogy with the bulk Hilbert space $\mathcal{H}_V$ the surface Hilbert space can be conveniently written as an orthogonal sum in the form
\begin{equation}
\mathcal{H}_{Hor}=\bigoplus_{(\vec{P},b)}\mathcal{H}^{\vec{P},b}_{Hor}
\label{QG_002}
\end{equation}
where now $\vec{P}$ stands for an ordered $n$-tuple of points on the ``horizon'' $S$ labeled by integers mod $\kappa$ ($b_i\in \mathbb{Z}_\kappa,\,i=1,\ldots,n$) satisfying the condition $b_1+\cdots+b_n=0$. { Here $\mathcal{H}^\emptyset=\{\vec{0}\}$.}

At this stage in the process both spaces are completely independent. { The key element that establishes a relationship between them is} the quantized version of the horizon boundary conditions that we have discussed at the end of the preceding section (\ref{U1Conds},\ref{SU2Conds}). It is very important to highlight here the fact that the operators that appear in these quantized boundary conditions are defined in completely unrelated Hilbert spaces; hence, the fact that there exist solutions to these quantum boundary conditions is highly  non-trivial. Of course, one has also to take into account the quantized constraints in the bulk Hilbert space by using the standard LQG methods (Dirac quantization, group averaging, etc., see {\bf Chapter LOOP QUANTUM  GRAVITY}). The implementation of the quantum boundary conditions leads to a subspace consisting of orthogonal sums of elements of the form
\begin{equation}
\mathcal{H}_{Bulk}
^{P,j,m}\otimes \mathcal{H}^{\vec{P},b}
\label{QG_003}
\end{equation}
such that the points in the set $P$ coincide with those in the vector $\vec{P}$ and the $b_i$ labels associated with the punctures satisfy the condition $b_i=-2m_i (mod,\kappa)$ for $i=1,\ldots,n$.

Up to this point the construction has given us some kind of kinematical Hilbert space adapted to the present situation where we have inner spacetime boundaries\footnote{Notice, however, that the quantum boundary conditions, arising from consistency requirements for the Hamiltonian formulation of the sector of general relativity that we are considering here, can also be thought of as constraints and, from this perspective, what we have really done is to implement them \textit{\`a la Dirac}.} We still have to take into account the rest of the constraints in the model. This is done by following the standard procedure { (see {\bf Chapter LOOP QUANTUM  GRAVITY})} and making some assumptions --presumed mild-- regarding solutions to the Hamiltonian constraint \cite{ABK}.

One of the key insights in the development of the present framework was the introduction by Krasnov \cite{Krasnov:1996tb,Krasnov:1996wc,Krasnov:1997yt} of the \textit{area ensemble}. In the absence of a suitable notion of energy such definition seemed natural: the area is an extensive quantity with a well understood discrete spectrum. This state of affairs has evolved due to results\cite{Frodden:2011eb} that provide an interpretation of the horizon area as a quasilocal notion of energy. This will be discussed in the last two sections of this chapter.  It is important to highlight, at this point, that area and angular momentum play a fundamental role already at the classical level in the IH framework whereas mass is a derived physical magnitude.

The customary way to define the entropy starts by considering the prequantized value of the area $A_\kappa$ and introducing an area interval $[A_\kappa-\delta, A_\kappa+\delta]$ of width $\delta$ of the order of the Planck length\footnote{A different construction is possible if one uses the so called flux operator \cite{nosLew} to define the entropy. { In this case there is no need to introduce an area interval to solve the quantum matching conditions though, on physical grounds, it is useful to introduce it afterwards.}}. Once this is done the entropy can be {computed} by tracing out the bulk degrees of freedom to define a density matrix describing a maximal mixture of states on the horizon surface $S$ with area eigenvalues in the previous interval. In order to count the number of states in $[A_\kappa-\delta, A_\kappa+\delta]$ we have to find out how many lists of non-zero elements of $\mathbb{Z}_\kappa$ satisfy the condition $\sum_{i=1}^n b_\kappa=0$ with $b_i=-2m_i (mod\,\kappa)$ for a permissible list of labels $m_1,\ldots$ By permissible we mean that there must exist a list of non-vanishing spin labels $j_i$ such that each $m_i$ is a spin component of $j_i$ ($m_i\in\{-j_i,-j_i+1\ldots, j_i\}$) and the following inequality holds
\begin{equation}
A_\kappa-\delta\leq 8\pi\gamma\ell_{Pl}^2\sum_{i=1}^n\sqrt{j_i(j_i+1)}\leq A_\kappa+\delta\,.
\label{QG_004}
\end{equation}
In principle the preceding discussion gives a concrete prescription that defines the counting (combinatorial) problem that has to be solved in order to compute the entropy for a given value of the prequantized area $A_\kappa$. This is generalized to arbitrary values of the area by allowing $A_\kappa$ to be replaced by any arbitrary value $A$.

The preceding combinatorial problem can be considered as is (and, in fact, when the flux operator is used it can be solved in a relatively straightforward way). However, there is a neat way to simplify it  know as the Domagala-Lewandowski (DL) approach \cite{Domagala:2004jt}. By carefully considering the details of the problem it is possible to pose it in such a way that only one type of labels appear (instead of the three labels in the original formulation, viz. $j_i,m_i,b_i$). { In the new rephrasing the entropy is computed as} $\log n(A)$ where $n(A)$ is 1 plus the number of finite sequences of non-zero integers or half-integers satisfying the following two conditions

\begin{equation}
\sum_{i=1}^n\sqrt{|m_i|(|m_i|+1)}\leq \frac{A}{8\pi\gamma \ell_{Pl}^2}\,,
\label{QG_005}
\end{equation}
and the so called projection constraint
\be
 \quad \sum_{i=1}^nm_i=0\,.
\ee
{  
A different approach corresponds to the models described by Ghosh and Mitra (GM) \cite{Ghosh:2004rq,Ghosh:2004wq} leading to the definition the entropy as $\log n(A)$ where $n(A)$ is 1 plus the number of all finite, arbitrarily long sequences $((j_1,m_1),\ldots,(j_N,m_N))$ of ordered pairs of non-zero, positive half integers $j_i$ and spin components $m_i\in\{-j_i,-j_i+1,\ldots,j_i\}$ satisfying
\begin{equation}
\sum_{i=1}^n\sqrt{j_i(j_i+1)}\leq \frac{A}{8\pi\gamma \ell_{Pl}^2}\,, \quad \sum_{i=1}^nm_i=0\,.
\label{QG_006}
\end{equation}

The difference between the DL and the GM definition of the counting problem resides in the following technical point.
As two punctures with different spins $j\not=j'$ but with the same magnetic number $m$ are, from the boundary $U(1)$ Chern-Simons theory, indistinguishable, they are considered as physically equivalent in the DL prescription.  In the GM prescription the previous two configurations are considered as different and counted individually. This apparent ambiguity of prescriptions disappears in the $SU(2)$ invariant formulation where, roughly speaking, the states of the Chern-Simons boundary connection depend both $j$ and $m$. To leading order the counting in the $SU(2)$ invariant formulation agrees with the GM prescription (see table in Section \ref{IH_counting}).

Up to this point we have described the $U(1)$ framework, which among other things, involves the quantisation of condition (\ref{U1Conds}) and its variants. Let us now briefly present the $SU(2)$ framework following from the quantisation of the system containing (\ref{SU2Conds}). The first models using $SU(2)$ Chern-Simons theory were proposed by Kaul and Majumdar \cite{Kaul:1998xv}.
The complete $SU(2)$ framework, including the classical description of the theory,  was proposed by Engle, Noui and Perez in \cite{Engle:2009vc,Engle:2010kt,Engle:2011vf}. The entropy in this case is computed as $\log n(A)$ where $n(A)$ is 1 plus the number of all finite, arbitrarily long sequences $(j_1,\ldots,j_N)$ of non-zero, positive half integers $j_i$ satisfying the inequality
\begin{equation}
\sum_{i=1}^n\sqrt{j_i(j_i+1)}\leq \frac{A}{8\pi\gamma \ell_{Pl}^2}\,,
\label{QG_007}
\end{equation}
and counted with multiplicity given by the dimension of the invariant subspace $\mathrm{Inv} \otimes_i[j_i]$.

The next section will be devoted to introducing efficient methods to solve the different types of combinatorial problems involved in the computation of the entropy in the different proposals. These methods are based in number-theoretic ideas and provide a powerful setup to deal with the broad class of problems arising in the study of black holes in LQG.} 


\section{Counting and number theory}\label{IH_counting}

The different models for semiclassical black holes described in the preceding { section} provide concrete { examples} of the kind of counting problems that have to be solved in order to compute the black hole entropy as a function of the area (and other physical features such as angular momentum). { They are remarkable} for several reasons. First, they are relatively easy to state and, in fact, reduce to the counting of specific types of finite sequences of integers or half integers subject to simple conditions. Furthermore their resolution can be tackled by using methods that combine known types of Diophantine equations, the use of generating functions and Laplace transforms.

As we will show in the following all the black hole models that have been discussed so far in the LQG framework lead to the Bekenstein-Hawking law. Some of them, in particular the older ones \cite{Ashtekar:1997yu,Kaul:1998xv,Ghosh:2006ph}, require the fine tuning of the Immirzi parameter to get the correct proportionality factor between area and entropy; others give results consistent with the first law and equation (\ref{entro0})  without any fine tuning  \cite{Ghosh:2011fc, Ghosh:2013iwa} (see Section \ref{semi}).  It is important to point out at this point that the fact that the entropy grows linearly in the asymptotic limit of large areas is not a \textit{generic} behavior. At first sight the situation seems to be quite similar to that of a sufficiently regular real function $f(A)$ satisfying $f(0)=0$ and $f'(0)\neq0$ for which Taylor's theorem implies that in the $A\rightarrow0$ asymptotic limit $f(A)\propto A$. However the limit that we are considering is $A \rightarrow\infty$ and the function of interest (the entropy) is not analytic but, actually, has a staircase form (though it can be written in terms of non-trivial integral expressions). In such circumstances the linear asymptotic behavior for large areas is certainly significant and becomes a { genuine nontrivial prediction} of the model.

We want to make some additional comments regarding the counting entropy before describing in some detail the mathematical methods necessary to efficiently solve the combinatorial problems involved in its computation. The first has to do with its behavior for \textit{small areas} that was considered in detail by Corichi, Diaz Polo and Fernandez Borja \cite{Corichi:2006bs,Corichi:2006wn}. Quite unexpectedly one finds a regular step structure that persists for a reasonably wide interval---microscopic in any case---of areas (a detailed account of the mathematical reasons for this phenomenon can be found in \cite{BV1}). This is mildly reminiscent of the predictions by Bekenstein, Mukhanov and others \cite{Bek1,Bek2} regarding a ``quantized'' area spectrum. In the face of it this does not seem to be utterly unexpected because the area operator, with its discrete spectrum, plays a central role in the formalism. However, the eigenvalues of the area are not equally spaced and their density (as a function of the area) grows very fast whereas the width of the steps seen in the entropy is both exact and persistent (although they eventually disappear).

A second relevant comment has to do with the rigorous notion of thermodynamic limit\cite{Griffiths}. This has important implications for the mathematical properties of the entropy as a function of its natural variables (the energy in the case of statistical mechanics). In this limit (that can be computed by working with the counting entropy that we are considering here) the entropy is smooth almost everywhere --which implies that standard thermodynamical formulae can be used-- and is \textit{concave} (downwards). A consequence of this last fact is that the step structure for small areas should not be directly observable (although it can possibly have some kind of impact on its properties). Another important consequence of this is the change in the predictions for the subdominant corrections to the entropy for large areas (that actually disappear for some models \cite{Engle:2009vc,Engle:2010kt,Engle:2011vf}).

The general structure of the combinatorial problems that have to be solved is the following. In all the cases one must count the number of finite, arbitrarily long, sequences of non-zero half integers satisfying an inequality condition involving the horizon area. These numbers are associated with spin network edges that pierce the horizon and quantum numbers coming from the Chern-Simons sector at the horizon. In the case of the original $U(1)$ proposal of Ashtekar, Baez, Corichi and Krasnov \cite{Ashtekar:1997yu} the associated combinatorial problem was rephrased in a convenient simplified way \cite{Domagala:2004jt,Meissner:2004ju} that did not involve directly the spin labels $j_i$ associated with the punctures at the horizon but, rather, the magnetic quantum numbers $m_i$ (satisfying the condition $-j_i\le m_i\le j_i$). For a given value of the horizon area these numbers have to satisfy the inequality
\begin{equation}
\sum_{i=1}^N\sqrt{|m_i|(|m_i|+1)}\leq \frac{A}{8\pi\gamma\ell_{Pl}^2}\,,\label{cond1}
\end{equation}
and { the \textit{projection constraint}}
\begin{equation}
\sum_{i=1}^N m_i=0\,,\label{cond2}
\end{equation}
In other proposals, such as { the GM prescription\cite{Ghosh:2006ph}}, the combinatorial problem is expressed in terms of both the spin labels $j_i$ of the edges that pierce the horizon and the $m_i$ labels. There is an inequality (similar to \ref{cond1}) and a projection constraint with the same form as before
\begin{equation}
\sum_{i=1}^N\sqrt{j_i(j_i+1)}\leq \frac{A_k}{8\pi\gamma\ell_{Pl}^2}\,,\quad \sum_{i=1}^N m_i=0\,.\label{cond3}
\end{equation}
{ Notice how these two counting problems are different: in the first one both conditions involve the $m_i$ labels whereas in the second the $j_i$ and $m_i$ labels are quite independent (though they must satisfy the restriction $-j_i\le m_i\le j_i$). In the $SU(2)$ models\cite{Engle:2009vc,Engle:2010kt,Engle:2011vf}} the projection constraint is replaced by a condition involving the dimension of the invariant subspace. Lack of space precludes us from delving into the details of all the different cases and proposals so we will describe only the DL approach in some detail and refer the reader to the literature { for the rest}. In the rest of this section we will use units such that $4\pi\gamma\ell_{Pl}^2=1$.

In order to count the number of sequences as required by the previous prescription it is convenient to adopt a stepwise approach. This has been explained in detail elsewhere \cite{Agullo:2008yv,Agullo:2010zz} so we give here a summary of the procedure. The main steps are:

\begin{enumerate}
\item[1.] For each fixed value of the area $a$ obtain all the possible choices for the positive half integers $|m_i|$ { \textit{compatible}} with it in the sense that they satisfy
    \begin{equation}
    \sum_{i=1}^N\sqrt{|m_i|(|m_i|+1)}= \frac{A}{2}.
    \label{condicion_1}
    \end{equation}
    At this stage the numbers $|m_i|$ can repeat themselves and are not ordered. In other words, in this first step we just want to find out how many times each spin component appears (how many $1/2$'s, how many $1$'s, and so on).
\item[2.] Count the different ways in which the multiset just described  can be reordered.
\item[3.] Count all the different ways of introducing signs in the sequences of the previous step in such a way that the condition $\sum_{i=1}^N m_i=0$ is satisfied.
\item[4.] Repeat this procedure for all the eigenvalues of the area operator smaller than $A$ and add up the number of sequences thus obtained.
\end{enumerate}
The first step is a characterization of the part of the spectrum of the area operator relevant to the computation of black hole entropy, in particular the degeneracy of the area eigenvalues. The condition (\ref{condicion_1}) can be rewritten as
\begin{eqnarray}
\sum_{k=1}^{k_{\rm max}}N_k\sqrt{(k+1)^2-1}=A\hspace{9mm}\label{eqkn_k}
\end{eqnarray}
where we have introduced integer labels $k_i:=2|m_i|$. The non-negative integers $N_k$ (that will be allowed to be zero) in the last sum tell us the number of times that the label $k/2\in\mathbb{N}/2$ appears in the sequence. We also denote as $k_{\rm max}=k_{\rm max}(A)$ the maximum value of the positive integer $k$ compatible with the givem area $A$. The problem that we need to solve at this step can be rephrased as that of finding all the sets of pairs $\{(k,N_k):k\in\mathbb{N},\,N_k\in\mathbb{N}\cup\{0\}\}$ satisfying (\ref{eqkn_k}). It is important to notice now that (\ref{eqkn_k}) implies that the area eigenvalue $a$ must be an integer linear combination of square roots of squarefree numbers of the form $A=\sum_{i=1}^{i_{\rm max}} q_i \sqrt{p_i},\quad q_i\in\mathbb{N}\cup\{0\}\,,$ so that we have the condition
\begin{equation}
\sum_{k=1}^{k_{\rm max}}N_k\sqrt{(k+1)^2-1}=\sum_{i=1}^{i_{\rm max}} q_i\sqrt{p_i}\,.
\label{eqdiof1}
\end{equation}
where the right hand side is fixed from the initial choice of area eigenvalue $A$. The resolution of the previous equation is quite direct although the procedure, that involves the solution of the quadratic Diophantine equation known as the Pell equation and an auxiliary set of linear Diophantine equations, is somewhat lengthy. The interested reader is referred to \cite{Agullo:2008yv,Agullo:2010zz} for details. The final result of the analysis sketched at this step is a characterization of the number of times that a each spin label corresponding to a puncture can appear for a given area eigenvalue.

The second step simply requires us to count the number of ordered sequences containing the number of each label obtained in the previous step and is completely straightforward. Once we have found all the possible sequences of positive half-integers $|m_i|$ satisfying  condition \ref{condicion_1}, the third step asks for the computation of the number of ways to introduce \textit{signs} in each $m_i$ in such a way that the condition
$\sum_{i=1}^N m_i=0$ is satisfied. There are several ways to solve this problem as described in \cite{Agullo:2010zz}. The simplest one makes use of generating functions and is actually the preferred one as generating functions play a fundamental role in this framework { (as first explored and explained by Hanno Sahlmann\cite{Sahlmann:2007jt,Sahlmann:2007zp})}. The other methods are interesting because they suggest deep connections between the ideas presented here with other physical problems, in particular those involving conformal field theories \cite{Agullo:2009zt}.

The last step requires us to add up the number of configurations corresponding to all the area eigenvalues smaller or equal than $A$. The best way to do this makes again use of generating functions \cite{Sahlmann:2007jt,Sahlmann:2007zp,Agullo:2008yv} and Laplace transforms (see { references} \cite{Meissner:2004ju,Agullo:2008yv}). Generating functions are a very powerful tool in combinatorics because they can encode a lot of useful information about a particular problem and can be manipulated with very simple analytical tools. In the present case the step by step procedure
described above leads to concrete forms for the generating functions for all the problems described before and others considered in the literature \cite{Engle:2009vc,Engle:2010kt,Engle:2011vf}. In the specific example of the DL counting the generating { function is}\cite{BarberoG.:2008ue}
\begin{eqnarray}
G^{\rm DL}(z,x_1,x_2,\dots)&=&\left(\displaystyle 1-\sum_{i=1}^\infty\sum_{\alpha=1}^\infty
\Big( z^{k^i_\alpha}+z^{-k_\alpha^i}\Big) x_i^{y^i_\alpha}\right)^{-1}\,.\label{GDL}
\end{eqnarray}
Here the pairs $(k^i_\alpha,y^i_\alpha)$ are solutions { to} the Pell equation defined by the $i$-th square free integer. The coefficients $[z^0][x_1^{q_1}x_2^{q_2}\cdots]G^{\rm DL}(z,x_1,x_2,\dots)$ contain the information on the number of configurations compatible with a certain value of the area $\sum q_i\sqrt{p_i}$ . Once the generating function is at hand it is possible to use it to get a very useful integral representation \cite{Meissner:2004ju,G.:2008mj} that takes the form of an double inverse Laplace-Fourier transform.

The usefulness of Laplace transforms to deal with counting problems in this setting was pointed out by Meissner in reference\cite{Meissner:2004ju}. In addition to providing a way to effectively deal with step 4 in our scheme it is important also from the point of view of statistical mechanics and has been used to gain some understanding about the thermodynamic limit for black holes \cite{G.:2011zr}. The underlying reason is the fact that the passage from the microcanonical to the canonical ensembles can be understood \textit{precisely} in terms of Laplace transforms.
This way we get the following expression for the entropy
\begin{eqnarray}
\exp S(A)&=&\label{numint}\\
&&\frac{1}{(2\pi)^2 i}\int_{0}^{2\pi} \int_{x_0-i\infty}^{x_0+i\infty}
s^{-1}\Big(\displaystyle 1-2\sum_{k=1}^\infty e^{-s\sqrt{k(k+2)}}\cos \omega k\Big)^{-1} e^{As}\,\mathrm{d}s \,\mathrm{d}\omega\,,\nonumber
\end{eqnarray}
where $x_0$ is a real number larger than the real part of all the singularities of the integrand.\footnote{This expression is actually valid only for those values of the area $a\geq0$ \textit{that do not belong to the spectrum of area operator} whereas for $a_n$ in the spectrum of the area operator it gives the arithmetic mean of the left and right limits when $a\rightarrow a_n^{\pm}$. In practice the integral representation contains all the information about the entropy in a useful form.}
The treatment of other models such as the ones proposed by Ghosh, Mitra (GM), and Engle, Noui, Perez (ENP)  basically differ only on the treatment of the projection constraint. Relevant details can be found in reference\cite{Agullo:2010zz}.

In addition to the cases mentioned above it is sometimes useful to consider the simplified model in which the projection constraint is ignored. Physically this corresponds to a situation in which the entropy satisfies the Bekenstein-Hawking law with no logarithmic corrections. In this simplified example the generating function is just
\begin{eqnarray}
G_{DL(0)}(x_1,x_2,\dots)&=&\left(\displaystyle 1-2\sum_{i=1}^\infty\sum_{\alpha=1}^\infty
x_i^{y^i_\alpha}\right)^{-1}\,.\label{GDL_noproy}
\end{eqnarray}
leading to the following expression for the entropy
\begin{equation}
\exp S(A)=\frac{1}{2\pi i}\int_{x_0-i\infty}^{x_0+i\infty}
s^{-1}\Big(\displaystyle 1-2\sum_{k=1}^\infty e^{-s\sqrt{k(k+2)}}\Big)^{-1} e^{A s}\,\mathrm{d}s \, .\label{entro_noproy}
\end{equation}

Let us now briefly explain how the asymptotic behavior of the entropy is obtained. To this end one should remember that whenever a function is represented as an inverse Laplace transform (a so called \textit{Bromwich integral} such as \eqref{entro_noproy}) its asymptotic behavior as a function of the independent variable (the area $A$ in this case) is determined by the analytic structure of the integrand, specifically the position of the singularity $s_0$ with the largest real part. In the present case, after reintroducing units for the sake of the argument, we get
\begin{equation}
S(A)=S_0+\frac{\mathrm{Re}(s_0)}{\pi\gamma}\frac{A}{4\ell_{Pl}^2}\,.
\label{BHlaw}
\end{equation}
where $S_0$ is a constant independent of the area $A$. The preceding expression tells us that we exactly recover the Bekenstein-Hawking law by choosing $\gamma$ such that
\begin{equation}
\gamma=\frac{\mathrm{Re}(s_0)}{\pi}\,.
\label{immirzi}
\end{equation}
The analytic structure of the integrand in \eqref{entro_noproy} has some very interesting features such as the accumulation of the real parts of its singularities \cite{G.:2008mj} that reflect themselves in the behavior of the entropy. The expression (\ref{entro_noproy}) is very useful to explore these issues as only the complex variable $s$ is relevant (the discussion when the projection constraint is also taken into account is slightly more involved).

In the present example the entropy $S(A)$ displays a simple linear growth for large values of the area (without any logarithmic corrections) with a slope that depends on the Immirzi parameter $\gamma$. An interesting fact is that the value of the parameter $\gamma$ is the same for a number of different types of black holes \cite{} although different proposals such as \cite{Ashtekar:1997yu,Ghosh:2004rq,Engle:2009vc,Engle:2010kt,Engle:2011vf} lead to different values for it. The larger ones correspond to those cases where the number of microstates is larger as can be easily deduced from \eqref{BHlaw}. At variance with this behavior it is interesting to mention that the subdominant logarithmic corrections are independent of $\gamma$.

The values of the Immirzi parameter leading to the { Bekenstein-Hawking law} and the logarithmic corrections for the different models and proposals are the following 
\begin{center}
\begin{tabular}{|c|c|c|c|}
  \hline
  \it{Approach} \T\B& $ \gamma$ \T\B&\it{ Log correction}  \T\B&\it{ { Log corr. therm. limit}} \\  \hline  \hline
  DL(0) & $\,\,\gamma_{\rm DL}=0.237\cdots$ & $0$ &{$\log(A/\ell_{Pl}^2)$} \\ \hline
  DL & $\,\,\gamma_{\rm DL}=0.237\cdots$ & $-\frac{1}{2}\log(A/\ell_{Pl}^2)$ &{$\frac{1}{2}\log(A/\ell_{Pl}^2)$}\\  \hline
  GM &  $\,\,\gamma_{\rm GM}=0.274\cdots$ & $-\frac{1}{2}\log(A/\ell_{Pl}^2)$ & \it{exercise}\\  \hline
  ENP &  $\gamma_{\rm ENP}=\gamma_{\rm GM}$  \phantom{....}& $-\frac{3}{2}\log(A/\ell_{Pl}^2)$ &{0}\\  \hline\hline
\end{tabular}
\end{center}
The first column refers to the four different models considered (the one provided by the DL prescription without and with the projection constraint, the GM approach and the ENP model). The difference between the results for the DL prescription with and without the projection constraint is the presence of a negative logarithmic correction for the latter. This is to be expected as the incorporation of the projection constraints eliminates some microstates that are taken into account in the DL(0) model. A similar argument applies to the GM and ENP cases; the $-3/2$ coefficient for the logarithmic correction in the latter case means that the number of microstates allowed is smaller than for the GM proposal. See the discussion at the end of Section \ref{IH_quantum geometry}.


\section{Semiclassical advances}\label{semi}

The indeterminacy, mentioned in Section \ref{IH}, of the quantities appearing in the first law for IHs disappears if one changes the point of view and  assumes that the near horizon geometry corresponds to that of a stationary black hole solution and shifts the perspective to that of a suitable family of stationary nearby local observers. As explained in Section \ref{IH} the whole idea behind isolated horizons is to describe a sector of the phase space of gravity containing a boundary with the geometric properties of a BH horizon in equilibrium and  infinitely many bulk degrees of freedom. In such context no condition in the definition requires the near-horizon geometry to be that of a stationary black hole. A key point is that the systems that behave thermodynamically are those solutions in the phase space of IH whose {\em near horizon geometry} (NHG) is that of a stationary black hole solution\cite{Frodden:2011eb, Ghosh:2011fc, Frodden:2011zz, Ghosh:2013iwa}.   In the quantum theory this would amount to selecting a bulk quantum state that is semiclassical and peaked on the stationary black hole configuration near the isolated horizon. 

At present there is not enough control on the nature of the physical Hilbert space to be able to describe such states in detail (progress is reported in {\bf Chapter Quantum Hamiltonian}). Nevertheless, one can assume that such states exist and bring in their semiclassical properties into the analysis. This 
{\em semiclassical input} has led to interesting new insights into the black hole entropy calculation that we will briefly review here. This perspective opens a variety of new questions and tensions waiting to be resolved. We shall discuss them in the following Section.

Assume that the NHG to be isometric to that of a  Kerr-Newman BHs\footnote{Such assumption is physically reasonable due to the implications of the no-hair theorem.}.
A family of stationary observers $\mfs O$ located right outside the horizon at a small
proper distance $\ell\ll \sqrt{A}$ is defined by those following the integral
curves of the Killing vector field
\begin{eqnarray}
\chi=\xi+\Omega\,\psi=\partial_t+\Omega\,\partial_\phi,
\end{eqnarray}
where $\xi$  and $\psi$ are the Killing fields associated with the stationarity and axisymmetry of Kerr-Newman spacetime respectively, while $\Omega$ is the horizon angular velocity. 
%
The four-velocity of $\mfs O$ is given by
\begin{align}\label{keyy}
&u^a=\frac{\chi^a}{\|\chi\|}.
\end{align}
It follows from this that $\mfs O$ are uniformly accelerated with an acceleration $a=\ell^{-1}+o(\ell)$ in the normal direction. 
These observers are the unique stationary ones that coincide with the {\em locally non-rotating observers}\cite{Wald:1984rg} or ZAMOs\cite{Thorne:1986iy} as $\ell\to 0$. As a result, their angular momentum is not exactly zero, but $o(\ell)$. 
Thus $\mfs O$ are at rest with respect to the horizon which makes them the preferred observers for studying thermodynamical issues from 
a local perspective.


It is possible to show that the usual first law (\ref{1st}) translates into a much simpler relation among quasilocal physical quantities associated with $\mfs O$\cite{Frodden:2011eb}.  As long as the spacetime geometry is well approximated by the Kerr Newman BH geometry in the local outer region between the BH horizon and the world-sheet of local observers at proper distance $\ell$, and, in the leading order approximation for $\ell/\sqrt{A}\ll 1$, the following  local 
first law holds
\begin{eqnarray}\label{tutiri}
\delta E =\frac{\overline\kappa}{8\pi}\delta A,
\end{eqnarray}
where 
$\delta E=\int_W T_{\mu\nu} u^\mu dW^\nu=\|\chi\|^{-1} \int_W T_{\mu\nu} \chi^\mu dW^\nu$ represents the flow of energy across the world-sheet $W$ defined by the local observers, and 
$\overline \kappa\equiv\kappa/(\|\chi\|)$. The above result follows from the conservation law $\nabla^a (T_{ab} \chi^b)=0$ that allows one to write
$\delta E$ as the flux of $T_{ab} \chi^b$ across the horizon. This, in turn, can be related to changes in its area  using the optical 
Raychaudhuri equations\cite{Frodden:2011eb}.

Two important remarks are in order:
 First, there is no need to normalize the Killing generator $\chi$ in any particular way. The calculation leading to (\ref{tutiri}) is invariant under the rescaling $\chi\to \alpha \chi$ for $\alpha$ a non vanishing constant. This means that the argument is truly local  and should be valid for more general black holes with a Killing horizon that are not necessarily asymptotically flat. This rescaling invariance of the Killing generator corresponds precisely to the similar arbitrariness of the generators of IHs as described in Section (\ref{IH}). 
The fact that equation (\ref{tutiri}) does not depend on this ambiguity implies that  the local first law makes sense in the context of the IH phase space as long as one applies it to those solutions that are isometric to stationary black hole solutions in the thin layer of width $\ell$ outside the horizon. The semiclassical input is fully compatible with the notion of IHs. 

Second, the local surface gravity $\bar\kappa$ is universal $\bar\kappa=\ell^{-1}$ in its leading order behaviour for $\ell/\sqrt{A}\ll 1$. This is not surprising and simply reflects the fact that in the limit $\sqrt{A}\to \infty$ with $\ell$ held fixed the NHG in the thin layer outside the horizon becomes isometric to the corresponding thin slab of Minkowski spacetime outside a Rindler horizon: the quantity $\bar\kappa$ is the acceleration of the stationary observers in this regime. Therefore, the local surface gravity loses all memory of the macroscopic parameters that define the stationary black hole (see Section \ref{ensemble} for further discussion). This implies that, up to a constant which one sets to zero, equation (\ref{tutiri}) can be integrated, thus providing an effective notion of horizon energy 
\be E=\frac{A}{8\pi G_{N}\ell},\label{ene}\ee
where $G_N$ is Newton's constant. 
Such energy notion is precisely the one 
to be used in statistical mechanical considerations by local observers.   
Similar energy formulae have been obtained in the Hamiltonian formulation of general relativity
with boundary conditions imposing the presence of a stationary bifurcate horizon \cite{Carlip:1993sa}.
The area as the macroscopic variable defining the ensemble has been
always used  in the context of BH models in loop quantum gravity. The new aspect revealed by the previous equation is its physical interpretation as energy for the local observers.

The thermodynamical properties of quantum IHs satisfying the NHG condition can be described using standard statistical mechanical
methods with the effective Hamiltonian that follows from equation (\ref{ene}) and the LQG area spectrum (see {\bf Chapter LOOP QUANTUM GRAVITY}), namely
\be\label{ham}
\widehat H|j_1,j_2\cdots\rangle=\left(\gamma \frac{\ell^2_{Pl}}{2 G_N\ell}  \sum_{p} \sqrt{j_p (j_p+1)}\right)\  |j_1,j_2\cdots\rangle
\ee
where $j_p$ are positive half-integer spins of the $p$-th puncture and $\ell_{Pl}=\sqrt {G\hbar}$ is the fundamental Planck length associated with the gravitational coupling $G$ in the deep Planckian regime.
The analysis that follows can be performed in both the microcanonical ensemble or in the canonical ensemble; ensemble equivalence is granted
in this case because the system is simply given by a set of non interacting units with discrete energy levels.

\subsection{Pure quantum geometry calculation}\label{newly}

In this section we compute black hole entropy first in the microcanonical ensemble
following a simplified (physicist) version \cite{Ghosh:2008jc} of the rigorous detailed counting of the previous section.
As  the canonical ensemble
becomes available with the notion of Hamiltonian (\ref{ham}), we will also derive the results in the canonical ensemble framework. 
The treatment in terms of the grand canonical ensemble as well as the equivalence of the three ensembles has been shown\cite{Ghosh:2011fc}.

Denote by $s_j$ the number of punctures of the horizon labelled by the spin $j$.
Ignoring the closure constraint, and in the $SU(2)$ Chern-Simons formulation of quantum IHs,  the number of states
associated with a distribution of distinguishable punctures $\{s_j\}_{j=\frac{1}{2}}^\infty$ is
\be
n(\{s_j\})=\prod\limits_{j=\frac{1}{2}}^{\infty}\frac{N!}{s_j!}\,(2j+1)^{s_j},
\ee 
where $N\equiv\sum_j s_j$ is the total number of punctures. The leading term of the microcanonical entropy can be associated with
$S=\log(n(\{\bar s_j\}))$, where $\bar s_j$ are the solutions of the variational condition 
\be\label{vary}
\delta \log(n(\{\bar s_j\}))+2\pi \gamma_0\delta C_1(\{\bar s_j\})+\sigma C_2(\{\bar s_j\})=0
\ee
where $2\pi \gamma_0$ (the $2\pi$ factor is introduced for later convenience) and $\sigma$ are Lagrange multipliers for the constraints
\ba\label{c1c2}
C_1(\{\bar s_j\})&=& \sum_j \sqrt{j(j+1)} s_j-\frac{A}{8\pi\gamma\ell_{Pl}^2}=0, \n \\ C_2(\{\bar s_j\})&=&\sum_j s_j-N=0.
\ea
In words, $\bar s_j$ is the configuration maximazing $\log(n(\{s_j\}))$ for fixed macroscopic area $A$ and number of punctures $N$.
Notice that $C_1$ was not imposed in the treatment of Section \ref{IH_counting}\footnote{The physicist method of this section can be made precise using the counting techniques of Section \ref{IH_counting}. The counting with fixed $N$ is proposed as an exercise to the reader who is referred to \cite{FernandoBarbero:2011kb} for relevant equations.}. Ignoring $C_1$
amounts to setting the punctures chemical potential $\bar\mu=0$.  However, as we will show here, allowing for non vanishing chemical potential provides 
a whole new look at the question of the dependence of entropy on the Immirzi parameter. 

A simple calculation shows that the solution to the variational problem (\ref{vary}) is
\be\label{leading}
\frac{\bar s_j}{N}=(2j+1)\exp(- 2\pi\gamma_0 \sqrt{j(j+1)} -\sigma),  
\ee
from which it follows, by summing over $j$, that the Lagrange multipliers are not independent
\be\label{condition}
\exp\sigma(\gamma_0)=\sum_j (2j+1)\exp(- 2\pi \gamma_0 \sqrt{j(j+1)}).
\ee
It also follows from (\ref{leading}), and the evaluation of $S=\log(n(\{\bar s_j\}))$,  that
\be
S=\frac{\gamma_0}{\gamma} \frac{A}{4\ell_{Pl}^2}+\sigma(\gamma_0) N.
\ee
What is the value of the Lagrange multiplier $\gamma_0$? As in standard 
thermal systems the value of $\gamma_0$ is related to the temperature of the system.
Its value is fixed by the requirement that
\be\label{ttt}
\left.\frac{\partial S}{\partial E}\right|_N^{-1}=T=\frac{\hbar}{2\pi \ell},
\ee 
where $E$ is the energy measured by quasilocal observers (\ref{ene}) and the
last equality on the right is the condition that the temperature be the Unruh temperature (as measured by the same semiclassical observers).
The previous condition allows one to express the Lagrange multiplier $\gamma_0$ in terms of the (otherwise arbitrary) Immirzi parameter $\gamma$, $G$, and $G_N$, namely
\be\label{para} \gamma_0=\gamma \frac{G}{G_N},\ee and thus
\be\label{entra}
S=\frac{A}{4\hbar G_N}+\sigma(\gamma) N.
\ee
%
%
where
\be
\sigma(\gamma)= \log[\sum_j (2j+1) e^{-2 \pi \gamma\frac{G}{G_N} \nn }]
\ee
Notice that the first term in the entropy formula 
is given by the Bekenstein-Hawking area law with the low energy value of Newton constant $G_N$; in other words it does not depend explicitly on the fundamental 
Planck length $\ell_{Pl}$ appearing in the area spectrum. Even though this is to be expected as the Bekenstein-Hawking term is a semiclassical quantity, the above result sheds new light
on a long standing discussion in the community as to which is the value of Newton's constant that should go into the area spectrum. Due to quantum effects Newton's 
constant is expected to flow from the IR regime to the deep Planckian one. The Planckian value of the gravitational coupling should be defined in terms of the fundamental 
quantum of area predicted by LQG yet the low energy value should appear in the entropy formula. The semiclassical input that enters the derivation of the entropy 
through the assumption of (\ref{ene}) is the ingredient that bridges the two regimes.

Finally, punctures are associated with a chemical potential which is given by
\be\label{cp}
\bar\mu=-T \left.\frac{\partial S}{\partial N}\right|_E=-\frac{\ell_{Pl}^2}{2\pi\ell}\,\sigma(\gamma)
\ee
which depends on the fiducial length scale $\ell$ and the Immirzi parameter, and where one is again evaluating the equation at the Unruh
temperature $T=\hbar/(2\pi \ell)$.

The above derivation can be done in the framework of the canonical and grandcanonical ensembles.
From the technical perspective it would have been simpler to do it using one of those ensembles. In particular
basic formulae allow for the calculation of the energy fluctuations 
which at the Unruh temperature are such that  
$
{(\Delta E)^2}/{\langle E\rangle^2}=\sO(1/N).
$
The specific heat at $T_{\va U}$ is
$C=N\gamma_0^2d^2\sigma/d\gamma^2$
which is  positive. This implies that as a thermodynamic system the IH is locally
stable. The specific heat tends to zero in the large $\gamma$ limit for fixed
$N$ and diverges as $\hbar\to 0$. The three ensembles give equivalent results \cite{Ghosh:2011fc}.

\subsubsection{The thermodynamical vs. the geometric first law}

By simply computing the total differential of the entropy (\ref{entra}) one finds the thermodynamical first 
law
\be\label{localla}
\delta E=\frac{\bar\kappa \hbar }{2 \pi} \delta S+ \bar\mu \delta N
\ee
In order to find a relationship with the geometric first law (\ref{1st}), one needs to assume that the spacetime geometry corresponds to 
that of a stationary black hole (for which (\ref{1st}) applies). If one does so then one can show (by simply reverting the argument that took one from (\ref{1st}) to (\ref{tutiri}) \cite{Frodden:2011eb})
that (\ref{localla}) is equivalent to 
 \be\label{globalla}
 \delta M=\frac{\kappa \hbar }{2 \pi} \delta S+\Omega \delta J+\Phi \delta Q+ \mu \delta N,
 \ee
 where $\mu=-\ell_{Pl}^2 \kappa \sigma(\gamma)/(2\pi)$ (the redshifted version of $\bar\mu$). At first sight the previous equation does not look like (\ref{1st}).  However, it is immediate to check that the exotic chemical potential term in (\ref{globalla}) cancels the term proportional to the number of punctures in the entropy formula (\ref{entra}). For (\ref{localla}) this is due  to the equation of state (\ref{cp}); for (\ref{globalla}) this is due to the form of $\mu$. Therefore, the above balance equation is just exactly the same as (\ref{1st}). 
 The different versions of the first law are presented in Table \ref{one}. Notice that only those on the left column are to be interpreted thermodynamically. Assuming the validity of  semiclassical consistency discussed here for general  accelerated observers in arbitrary local neighbourhoods\cite{Jacobson:1995ab}, the emergence of general relativity directly from the statistical mechanics of the polymer like structures of LQG has been argued\cite{Smolin:2012ys}.

\begin{table}[ht!!!!!!!!]
\tbl{Different versions of balance equations. On the left column one has the results coming 
from quantum geometry involving a chemical potential term. The semiclassical input of the area effective Hamiltonian in the quantum geometry statistical
 mechanics calculation leads to results that are consistent with the geometry first laws shown on the right column.}
{\begin{tabular}{@{}cccc@{}} 
\colrule
 & &  &  \\
& Quantum Statistical  Mechanics & & Classical Einstein gravity   \\
& &  &  \\
 Local & {\normalsize $\delta E=\frac{\bar\kappa \hbar }{2 \pi} \delta S+ \bar\mu \delta N$ }& {\normalsize$\Longleftrightarrow$} & {\normalsize $\delta E=\frac{\bar\kappa}{8\pi} \delta A $} \\
& &  &  \\
&{\normalsize $\Updownarrow$}& &{\normalsize$\Updownarrow$}  \\
& &  &  \\
Global &{\normalsize$\delta M=\frac{\kappa \hbar }{2 \pi} \delta S+\Omega \delta J+\Phi \delta Q+ \mu \delta N$} & {\normalsize$\Longleftrightarrow$} & {\normalsize$\delta M=\frac{\kappa \hbar }{2 \pi} \delta A+\Omega \delta J+\Phi \delta Q$} \\ 
& &  &  \\
\botrule
\end{tabular}
}
\begin{tabnote}
$^{\text a}$Moving along horizontally in this table is a trivial identity; moving vertically requires the background 
geometry to be a stationary black hole solution.
\end{tabnote}
\label{one}
\end{table}

\subsubsection{Recovering the results of Section \ref{IH_quantum geometry}}

As we mentioned above the key difference with the counting of Section \ref{IH_quantum geometry}
is the imposition of the constraint $C_2$ in (\ref{c1c2}). One can therefore recover the results by simply setting the Lagrange
multiplier $\sigma=0$ from the onset of the calculation in Subsection \ref{newly}. What happens then is that 
equation (\ref{condition}) completely fixes $\gamma_0$ to the numerical value: in the present case $\gamma_0=0.274...$.
Equation (\ref{para})---which continues to hold---introduces a strong constraint  between 
fundamental constants; namely
\be\label{esa}
\gamma \frac{G}{G_N} = \gamma_0= 0.274...,
\ee
which corresponds to equation (\ref{immirzi}) with the identification $\gamma_0={\rm Re}(s_0)/\pi$.
The previous equation implies that $S=A/(4G_N \hbar)$.
Therefore, by declaring that the chemical potential of punctures vanishes $\bar \mu=0$ (equivalently $\sigma=0$)
 the  semiclassical consistency, equation (\ref{ttt}), is satisfied at the price of restricting the fundamental constants as above. 
It has been proposed that the previous equation, relating low energy $G_N$ with the fundamental couplings $G$ and $\gamma$, could be interpreted in the context of the renormalisation group flow\cite{Jacobson:2007uj}. However, due to the completely combinatorial way in which $\gamma_0$ arises (which does not make reference to any dynamical notion) it is so far unclear how such scenario could be realized.
The contribution of matter degrees of freedom (`vacuum fluctuations') to the degeneracy of the area spectrum has been neglected in the derivation leading to (\ref{esa}). 

\subsection{Matter and holography }\label{holi}

In the treatments mentioned so far punctures are distinguishable. Let us see here what indistinguishability would change.
Instead of the microcanonical ensemble, we use now the grand canonical ensemble as this will  considerably shorten the derivations 
(keep in mind that all ensembles are equivalent). Thus we start from the canonical partition function which for a system 
of non interactive punctures is $Q(\beta,N)=q(\beta)^N/N!$ where the $N!$ in the denominator is the Gibbs factor
that effectively enforces indistinguishability, and the one-puncture partition function $q(\beta)$ is given by  
\be\label{qqq}
q(\beta)=\sum_{j=\frac{1}{2}}^{\infty} d_j \exp(-\frac{\hbar \beta \gamma_0}{\ell} \nn ),
\ee
where $d_j$ is the degeneracy of the spin $j$ state (for instance $d_j=(2j+1)$ as in 
the $SU(2)$ Chern-Simons treatment). The grand canonical partition function is
\be
\sZ(\beta,z)=\sum_{N=1}^{\infty} \frac{z^Nq(\beta)^N}{N!}=\exp(zq(\beta)).
\ee
From the equations of state $E=-\partial_{\beta} \log(\sZ)$, and $N=z\partial_z\log(\sZ) $ one gets
\ba\label{eqsta}
\frac{A}{8\pi G_N \ell}&=&-z\partial_{\beta} q(\beta)\n \\
N &=& z q(\beta)=\log(\sZ).
\ea
In thermal equilibrium at the Unruh temperature one has $\beta=2\pi\ell\hbar^{-1}$ and the $\ell$ dependence disappears from the previous equations.
However, for $d_j$ that grow at most polynomially in $j$, the BH area predicted by the equation is just Planckian and the number of punctures $N$ of order one.
Therefore, indistinguishability with degeneracies $d_j$ of the kind we find in the pure geometry models is ruled out because it cannot predict 
semiclassical BH's.

An interesting perspective\cite{Ghosh:2013iwa} arises in the framework of quasilocal observers. If one only restricts to quantum geometry degrees of freedom then $d_j=2j+1$ in the 
$SU(2)$ ENP treatment or $d_j=1$ in the GM and DL models.
Now, from the local observers perspective, the quantum state of the system close to the horizon appears as a highly excited state at inverse temperature $\beta=2\pi \ell/\ell_{Pl}^2$. Of course this state looks like the vacuum state  for freely falling observers (at scales smaller than the size of the BH).  These two dual versions of the same physics tell us that the quantum state describing the near-horizon physics contains more than just pure quantum geometric excitations.  
Very general results from quantum field theory on curved spacetimes imply that the quasilocal observers close to the horizon would find that the number of degrees of freedom grows exponentially with the horizon area according to (see for instance \cite{'tHooft:1984re})
\be\label{holos}
D \propto \exp(\lambda A/(\hbar G_N)),
\ee
where $\lambda$ is an unspecified dimensionless constant that cannot be determined due to two related issues:
On the one hand UV divergences of standard QFT introduce regularization ambiguities affecting the value of $\lambda$; on the other hand,
the value of $\lambda$ depends on the number of species of fields considered.
For that reason, here we only assume the qualitative exponential growth and will prove below that the ambiguity in $\lambda$ is completely removed by non perturbative quantum gravity considerations.
%
%

From (\ref{holos}) $D[\{s_j\}]=\prod_j d_j$ with $d_j=\exp(\lambda 8\pi \gamma_0\nn)$. For simplicity lets take $\nn\approx j+1/2$. We also introduce 
two dimensionless variables $\delta_{\beta}$ and $\delta_{h}$ and write
$\beta=\beta_{\va U}(1+\delta_\beta)$---where $\beta_{U}=2\pi \ell/\hbar$---and $\lambda=(1-\delta_h)/4$.
A direct calculation of the geometric series that follows from (\ref{qqq}) yields
\be
q(\beta)
=\frac{\exp(-\pi\gamma_0\delta(\beta))}{\exp(\pi\gamma_0\delta(\beta))-1},
\ee
where $\delta(\beta)=\delta_h+\delta_{\beta}$. The equations of state now predict large semiclassical 
BHs: for large $A/(\hbar G_N)$ equation (\ref{eqsta}) can be used to determine $\delta$ as a function of $A$ and $z$. The result is
$\delta=2\sqrt{{G_N \hbar z}/({\pi\gamma_0 A})}\ll 1$. For semiclassical BHs $\delta_{\beta}\ll 1$  since the temperature measured by quasilocal observers must be close to the Unruh temperature. This, together with the previous equation for $\delta$, implies $\delta_h\ll1$.
In other words semiclassical consistency implies that the additional degrees of freedom producing the degeneracy (\ref{holos}) must saturate the {\em holographic bound}  \cite{Ghosh:2013iwa}, i.e. $\lambda=1/4$ up to quantum corrections.

The entropy is given by the formula $S=\beta E-\log(z) N+\log(\sZ)$ which upon evaluation yields
\be
S=\frac{A}{4G_N\hbar} - \frac{1}{2}(\log(z)-1) \left({\frac{zA}{\pi \gamma \ell_{Pl}^2}}\right)^{\frac{1}{2}}
\ee
This gives the Bekenstein-Hawking entropy to leading order plus 
quantum corrections. If one sets the chemical potential of the punctures to zero (as for photons or gravitons)
then these corrections remain. One can get rid of the corrections by setting the chemical potential 
$\mu=T_U$. Such possibility is intriguing, yet the physical meaning of such a choice is not clear at this stage. 
The thermal state of the system is dominated by large spins as the mean spin $\langle j\rangle = A/(N\ell_{Pl}^2)$ grows like $\sqrt{A/\ell_{Pl}^2}$.
The conclusions of this subsection hold for arbitrary puncture statistics. This is to be expected because the system behaves as if it were at a very high effective temperature  (the Unruh temperature is the precise analog of 
the Hagedorn temperature of particle physics) \cite{Ghosh:2013iwa}. Because it will be important for further discussion we write the partition function
corresponding to the choice of Bosonic statistics of punctures explicitly and for $z=1$, namely
\be\label{z=1}
\sZ(\beta)=\prod_{j=\frac{1}{2}}^\infty \sum_{s_j} \exp(2\pi\ell-\beta)\frac{a_j}{8\pi\ell G_N},
\ee
where $a_j=8\pi\gamma\ell_{Pl}^2\nn$ are the area eigenvalues, and we have assumed for simplicity $\lambda=1/4$ in (\ref{holos}), namely $d_j=\exp(a_j/(4G_N\hbar))$.
Interestingly, such exact holographic behaviour of the degeneracy of the area spectrum can be obtained from an analytic continuation of the 
dimension of the boundary Chern-Simons theory by sending the spins $j_i\to i s-1/2$  with $s\in \R^+$\cite{Frodden:2012dq, Han:2014xna, Achour:2014eqa, Geiller:2013pya}. 
The new continuous labels correspond to $SU(1,1)$ unitary representations that solve the $SL(2,\C)$ self(antiself)-duality constraints 
$L^i\pm K^i=0$ (see {\bf Chapter SPIN FOAMS}), which in addition comply with the necessary reality condition $E\cdot E\ge 0$ for the fields $E^a_i$ (see {\bf Chapter LOOP QUANTUM GRAVITY}).
All this suggests that the holographic behaviour postulated in (\ref{holos}) with $\lambda=1/4$ would naturally follow from the definition of LQG in terms of self(antiself)-dual variables, i.e.
$\gamma=\pm i$.  The same holographic behaviour of the number of degrees of freedom available at the horizon surface
is found from a conformal field theoretical perspective for $\gamma=\pm i$ \cite{Ghosh:2014rra}. A relationship between the termal nature of BH horizons and self dual variables seems also valid according to similar analytic continuation arguments \cite{Pranzetti:2013lma}.    
The analytic continuation technique has also been applied in the context of lower dimensional BHs \cite{Frodden:2012nu}.

\subsubsection{What is the ensemble in the quasilocal treatment}\label{ensemble}

The quasi local perspective provides a description complementary  to the isolated horizon 
definition of the horizon Hilbert space. It allows one to perform manipulations in the canonical ensemble language.
At the basic level the ensemble is still defined by the details of the isolated horizon boundary conditions which tell us whether we are dealing with a spherical, distorted, rotating or static BH horizon. Even when charge and angular momentum do not appear in the expression of the quasi local first law these parameters (and all multipole moments in the case of distorted isolated horizons \cite{Ashtekar:2004gp, Ashtekar:2004nd}) are encoded implicitly in the form of the boundary condition used to define the quantum theory of the horizon.  Notice also that the usual canonical ensemble is ill defined\cite{Hawking:1976de} because the number of states grows too fast as a function of the ADM energy. This problem disappears in the quasi local treatment where the area ensemble plays the central role.


\section{Synergy as well as tension between the microscopic and
semi-classical descriptions}

\subsection{Spinfoams}\label{sf}

In the covariant path integral representation of loop quantum gravity
the state of a puncture (open spin network link) $|j,m\rangle$  is 
embedded in the unitary representations of $SL(2,\C)$ (whose basis vectors can be written as $|p, k; j,m\rangle$ for $p\in \R^+$ and $k\in \N$)
according to $|j,m\rangle\to |\gamma (j+1), j ; j,m\rangle$. The maximum weight states $m=j$ define a puncture state which is in turn a coherent state peaked 
along the $z${\em -axis} which is assumed (through an implicit gauge fixing) to correspond to the normal to the horizon. We denote such states as follows
\be
\mathbf { |j\rangle} \equiv |\gamma (j+1), j ; j,j\rangle.
\ee
These states satisfy the simplicity constraints $L^i=\gamma K^i$ in a weak sense (see {\bf Chapter SPIN FOAMS}).
One postulates\cite{Bianchi:2012vp} that quantum horizon states (in the infinite area limit, i.e. Rindler states) evolve in the {\em time} 
of stationary observers (\ref{keyy})---uniformly accelerated with $a=\ell^{-1}$---according to   
\be\label{euge}
\mathbf { |j_t\rangle}=\exp(i H t)  \mathbf { |j\rangle},
\ee
with $H= a K_z =\ell^{-1} K_z $ the Rindler Hamiltonian.
This time evolution is consistent with the semiclassical condition (\ref{ene}).  More precisely from the simplicity constraints one has that $\langle\mathbf j| H | \mathbf j\rangle=\hbar\gamma j \ell^{-1}$
which coincides with the eigenvalue of $E=A/(8\pi G_N \ell)$ for a single plaquette in the large $j$ limit. By coupling the system with an idealized detector modelled by a two level system
 \cite{Unruh:1976db} with energy separation $T_U=\hbar/(2\pi\ell)\ll\Delta \epsilon$ it is shown that
the population of the excited state in the stationary state is \cite{Bianchi:2012vp}
\be
p_1\approx \exp(-\frac{2\pi\ell}{\hbar} \Delta \epsilon ),
\ee
which is the Wien distribution at temperature $T_U=\hbar/(2\pi\ell)$. 
A key property\cite{Huszar:1971nn} leading to this result is the fact that
\be
|\langle \mathbf{\lambda} |\mathbf j\rangle|^2 \approx \lambda^{2j} \exp(-\pi \lambda) 
\ee
where $|\mathbf \lambda \rangle= |\gamma (j+1), j ; \lambda,j\rangle$ is an eigenstate of $K_z$ and $L_z$ with 
eigenvalues $\lambda$ and $j$ respectively.
In relation to this it has been postulated\cite{Chirco:2014saa} that
the one puncture reduced density matrix measuring the {\em inside-outside} correlations in spin foams is given by
\be
\rho_p=\frac{\exp(- 2\pi K_z)}{Z},
\ee
where $Z_p={\rm Tr}[\exp 2\pi K_z]$.
The single puncture entropy $S_p=-{\rm Tr}(\rho_p\log(\rho_p))=a_p/(4G\hbar)+\log(Z_p)$; by adding this result for all punctures
one gets a result of the form (\ref{entra}) with $Z_p$ playing the role of $\sigma$ (notice that both are the single puncture partition function).
In this way the results of the covariant and canonical approach are consistent. Notice that the fundamental input in the 
derivation of the temperature is that quantum horizon physical states are of the form (\ref{euge}). 

\subsection{Entanglement entropy perturbations and black hole entropy}

Starting from a pure state $|0\rangle\langle 0|$ (``vacuum state") one can define a reduced 
density matrix $\rho={\rm Tr}_{in}(|0\rangle\langle 0|)$ by taking the trace over the degrees of freedom
{\em inside} the BH horizon.  
The entanglement entropy is defined as $S_{ent}[\rho]=-{\rm Tr}(\rho\log(\rho))$.
In four dimensions\cite{Solodukhin:2011gn}  the leading order term of entanglement entropy in standard QFT goes like \be\label{entro} S_{ent}= \lambda \frac{A}{\epsilon^2}+corrections\ee 
where $\epsilon$ is an UV cut-off, and  $\lambda$ is left undetermined in the standard QFT calculation
due to UV divergences and associated ambiguities.  An important one is that 
$\lambda$ is proportional to the number of fields considered; this is known as the {\em species problem}. 
These ambiguities disappear if one studies perturbations of (\ref{entro}) 
when gravitational effects are taken into account \cite{Bianchi:2012br, Bianchi:2013rya}.
The analysis is done in the context of perturbations of the vacuum state in Minkowski spacetime
as seen by accelerated Rindler observers. Entanglement entropy is defined by tracing out degrees of freedom
outside the Rindler wedge. Such system reflects some of the physics of stationary black holes
in the infinite area limit.  A key property\cite{Wald:1995yp} is that, formally, 
\be\label{eeqq}
\rho=\frac{\exp (-2\pi \int_{\Sigma} \hat T_{\mu\nu}\chi^\mu d\Sigma^\nu)}{{\rm Tr}[\exp (-2\pi \int_{\Sigma} \hat T_{\mu\nu}\chi^\mu d\Sigma^\nu)]},
\ee
where $\Sigma$ is any Cauchy surface of the Rindler wedge.
If one considers a perturbation of the vaccum state $\delta \rho$
then the first interesting fact is that the (relative entropy) $\delta S_{ent}=S_{ent}[\rho+\delta\rho]-S_{ent}[\rho]$ is UV finite and hence free of regularization ambiguities \cite{Casini:2008cr}. 
The second fact is that due to (\ref{eeqq}) one has 
\ba \delta S_{ent}=2\pi {\rm Tr}(\int_{\Sigma}  \delta\langle T_{\mu\nu}\rangle\chi^\mu d\Sigma^\nu).\ea
Now from semiclassical Eintein's equations $\nabla^{\mu} \delta \langle T_{\mu\nu}\rangle=0$, this (together with the global properties of the Rindler wedge) implies that one can replace the Cauchy surface $\Sigma$ by the Rindler horizon $H$ in the  previous equation. As in the calculation leading to (\ref{tutiri}) one can use the Raychaudhuri equation (i.e. semiclassical Eintein's equations)  to relate the flux of $\delta \langle T_{\mu\nu}\rangle$ across the Rindler horizon to changes in its area. The result is that $\delta S_{ent}={\frac{ \delta A}{4G_N\hbar}}$ independently of the number of species.
The argument can be generalized to static black holes\cite{Perez:2014ura} where a preferred vacuum state exists (the Hartle-Hawking state). However, due to the fact that the BH horizon is no longer a good initial value surface the resulting balance equation is \be\label{dent}
\delta S_{ent}={\frac{ \delta A}{4G_N\hbar}}+\delta S_{\infty},
\ee
where $\delta S_{\infty}=\delta E/T_{H}$, and $\delta E$ is the energy flow at $\sI^+\cup i^+$.
Changes of entanglement entropy match changes of Hawking entropy plus an entropy flow to infinity.
These results shed light on the way the species problem could be resolved in quantum gravity. However,  as the concept of relative entropy used here is insensitive to the UV degrees of freedom,
the key question\cite{Perez:2014ura}
 is whether the present idea can be extrapolated to the Planck scale.
The results described in Section \ref{holi} go in this direction.

\subsection{Entanglement entropy vs. statistical mechanical entropy}\label{sf}

One can argue that the perspective that BH entropy should be accounted for in terms of entanglement entropy\cite{Solodukhin:2011gn}  and the
statistical mechanical derivation presented in this chapter are indeed  equivalent in a suitable sense\cite{Perez:2014ura}.  The basic reason for such equivalence resides in the microscopic structure predicted by LQG \cite{Bianchi:2012ui, Bianchi:2012ev, Chirco:2014saa}.  
In our context, the appearance of the UV divergence in (\ref{entro}) tells us that the leading contribution to $S_{ent}$ must come
from the UV structure of LQG close to the boundary separating the two regions. 
Consider a basis  of the subspace of the horizon Hilbert space characterised by condition (\ref{QG_004}), and assume the discrete index $a$ labels the elements of its basis. 
Consider the  state 
\be \label{here}
|\Psi\rangle=\sum\limits_{a} \alpha_{ a } \,  |\psi^{a}_{int}\rangle\, |\psi^{a}_{ext}\rangle, \ee 
where $|\psi^{a}_{int}\rangle$ and $|\psi^{a}_{ext}\rangle$  denote physical states compatible with the IH boundary data $a$, and describing  the interior and the exterior state 
of matter and geometry of the BH respectively. The assumption that such states exist is a basic input of Section \ref{IH_quantum geometry}. 
In the form of the  equation above we are assuming that correlations  between the outside and the inside at Planckian scales  are  mediated by the
spin-network links puncturing the separating boundary.
This encodes the idea that vacuum correlations are ultra-local at the Planck scale. 
This assumption is implicit in the recent treatments\cite{Bianchi:2012ui} based on the analysis of a single quantum of area correlation and it is related to the
(Planckian) Hadamard condition as defined in \cite{Chirco:2014saa}. 
We also assume states to be normalized as follows: $\langle \psi_{ext}^a|\psi_{ext}^a\rangle=1$, $\langle \psi_{int}^a|\psi_{int}^a\rangle=1$, and  $\langle \Psi|\Psi\rangle=1$.
The reduced density matrix obtain from the pure state by tracing over the interior observables is
\ba
 \rho_{ext}
&=&\sum\limits_a p_a |\psi^{a}_{ext}\rangle\langle \psi^{a}_{ext}|, 
\ea
with $p_a=|\alpha_a|^2$. It follows from this that the entropy $S_{ext}\equiv -{\rm Tr}[\rho_{ext} \log(\rho_{ext})]$ is  bounded by micro-canonical entropy of the ensemble (\ref{QG_004}) as discussed in Section \ref{IH_quantum geometry}. 
If instead one starts from a mixed state encoding an homogeneous statistical mixture of quantum states compatible with (\ref{QG_004}), then the reduced density matrix leads to an entropy that matches the microcanonical one\cite{Perez:2014ura}.

\subsection{Euclidean path integral (the quasi local treatment) and logarithmic corrections.}

Here we review some basic features of the Euclidean path integral approach to the computation of BH entropy. Although the method is formal, as far as the contribution of geometric degrees of freedom is concerned, it allows one to study the contributions of matter degrees of freedom in the vicinity of the horizon.  The formalism is relevant for discussing two important points. On the one hand it allows one to compare the partition function obtained in Section \ref{holi} with the field theoretical formal expression (providing in this way another test for semiclassical consistency). On the other hand it provides one with the tools that are necessary for comparison and discussion of the issue of logarithmic corrections in LQG and in other approaches. 

There is a well known relationship between the statistical mechanical partition function and the Euclidean path integral on a flat background.
One has that
\be
Z_{sc}(\beta)=\int D\phi \exp\{-S[\phi]\}
\ee
where field configurations are taken to be periodic in Euclidean time with period $\beta$.
Such expression can be formally extended  to the gravitational context at least in the treatment of stationary black holes.
One starts from the formal analog of the previous expression and immediately uses the stationary phase approximation to make sense of it on the background of a stationary black hole.
Namely 
\ba\label{EPI}
Z_{sc}(\beta)&=&\int Dg D\phi \exp\{-S[g, \phi]\}\n \\
&\approx& \exp\{-S[g_{cl}, 0]\} \int D\eta \exp\left[-\int dx dy \eta(x) \left(\frac{\delta^2 \sL}{\delta \eta(x)\delta \eta(y)}\right) \eta(y)\right]
\ea
where the first term depends entirely on the classical BH solution $g_{cl}$ 
while the second term represents the path integral over fluctuation fields, both of the metric as well as the matter, that we here schematically denote by $\eta$.  
For local field theories $\delta_{\eta(x)}\delta_{\eta(y)}{\sL}=\delta(x,y) \square_{gc}$ where $\square_{gc}$ is a the Laplace like operator 
(possible gauge symmetries, in particular diffeomorphisms must be gauge fixed to make sense of such formula).

Let us first concentrate on the evaluation of the classical action. In the quasi local treatment,  the Euclidean space time region, where the fields $\eta$ are supported, is given by a $D\times S^2$ where $D$ is a disk in a plane orthogonal to and centred at the horizon radius and having a proper radius $\ell$ (recall that in the Euclidean case the BH horizon shrinks to a point, represented here by the center of $D$). 
Using the Gibbons-Hawking prescription for the boundary term\cite{Gibbons:1976ue}, the action $S[g_{cl}, 0]$ is
\be
S[g_{cl}, 0]=\frac{1}{8\pi G_{N}} \left[ \ \int\limits_{D\times S^2} \!\!\!\!\!\sqrt{g} \, R+\int\limits_{\partial D\times S^2}\!\!\!\!\! (K-K_0)\, d\Sigma\right]
\ee
On shell the bulk term in the previous integral would vanish. However, unless $\beta_{H}=2\pi \kappa^{-1}$, the geometry has a conical singularity at the centre of the disk and the first term will contribute. The boundary term is the usual one with $K$ the extrinsic curvature of the boundary, $d\Sigma$ its volume form, and $K_0=-1/\ell$ is the value of the extrinsic curvature at the boundary in the $A\to \infty$ limit (Rindler space-time). The subtraction of the counter term $K_0$ has the same effect as replacing the inner conical singularity by an inner boundary with a boundary term of the form $\beta_{H} A/(8\pi)$ \cite{Carlip:1993sa}.  
A direct calculation gives the semi-classical free energy
\be\label{holographic}
-S[g_{cl}, 0]=\log(Z_{cl})=(2\pi \ell-\beta) \frac{A}{8\pi G_{N} \ell},
\ee
where $\beta=\beta_{H} \|\chi\|$ is the local energy. The equation of state $E=-\partial_{\beta} \log(Z_{cl})$ reproduces the quasilocal energy (\ref{ene})---this is a consequence 
of the substraction of $K_0$ \cite{York:1986it}.
The entropy is $S=\beta E+\log(Z)=A/(4\ell_{Pl}^2)$ when evaluated at the inverse Unruh temperature $\beta_U=2\pi\ell$.
Notice that in the quasi-local framework used here, entropy grows linearly with energy (instead of quadratically as in the usual 
Hawking treatment \cite{}). This means that the usual ill behaviour of the canonical ensemble of the standard 
global formulation\cite{Hawking:1976de} is cured by the quasilocal treatment. 

Notice that equation (\ref{holographic}) matches in form the partition function (\ref{z=1}).
In other words, the inclusion of the holographic degeneracy (\ref{holos}) plus the assumption of 
Bosonic statistics for punctures makes the results of section \ref{holi} compatible with the continuous 
formal treatment of the Euclidean path integral. In essence (\ref{z=1}) is a regularization of (\ref{EPI}).

Quantum corrections to the entropy come from the fluctuation factor 
which can formally be expressed  in terms of the determinant of a second order local (elliptic) differential operator $\square_{g_{cl}}$ 
\ba
&& F
=\int D\eta \exp\left[-\int dx \eta(x) \square_{g_{cl}} \eta(x)\right]=\left[{\det(\square_{g_{cl}})}\right]^{-\frac{1}{2}}.
\ea
The determinant can be computed from the identity (the heat kernel expansion)
\be
\log\left[{\det(\square_{g_{cl}})}\right]
=\int_{{\epsilon}^2}^{\infty} \frac{ds}{s} {\rm Tr}\left[\exp(-s \,\square_{g_{cl}})\right],
\ee
where $\epsilon$ is a UV cut-off needed to regularize the integral. We will assume here that it is proportional to $\ell_{Pl}$.
In the last equality we have used the heat kernel expansion in $d$ dimensions \be{\rm Tr}\left[\exp(-s\, \square_{g_{cl}})\right]=(4\pi s)^{-\frac{d}{2}} \sum_{n=0}^{\infty} a_n s^{\frac{n}{2}}, \ee where the coefficients $a_n$ are given by integrals in $D\times S^2$ of local quantities.  


At first sight the terms $a_n$ with $n\le 2$ produce potential important corrections to BH entropy. All of these suffer from regularisation ambiguities with the exception of the term $a_2$ which leads to logarithmic corrections.
Moreover, contributions coming from $a_0$ and $a_1$ can be shown to contribute to the renormalization of various couplings in the underlying Lagrangian \cite{Sen}; for instance $a_0$ contributes to the cosmological constant renormalization. 
True loop corrections are then encoded in the logarithmic term $a_2$ and for that reason it has received great attention in the literature (see \cite{Sen} and references therein). Another reason is that its form is regularisation independent.   
According to \cite{G.:2011zr} there are no logarithmic corrections in the $SU(2)$ pure geometric 
model once the appropriate smoothing is used (canonical ensemble). From this we conclude that the only possible source of logarithmic corrections in the $SU(2)$ case must come from the non-geometric degrees of freedom that produce the so called matter degeneracy that plays a central role in Section \ref{holi}. A possible way to compute these corrections is to compute the heat kernel coefficient $a_2$ for a given matter model. This is the approach taken in reference \cite{Sen}. 
One can argue\cite{Ghosh:2013iwa} that logarithmic corrections in the one-loop effective action are directly  reflected as logarithmic corrections in the LQG  BH entropy. The preceding considerations partially dissipate the perceived tensions between the LQG approach and others. This is an important question that deserves further attention. 

\subsection{Hawking radiation}\label{HR}

The derivation of Hawking radiation from first principles in LQG remains an open problem, this is partly due to the difficulty associated with the definition of semiclassical states approximating space-time backgrounds. Without a detailed account of the emission process it is still possible to obtain information from a spectroscopical approach that uses as an input the details of the area spectrum in addition to some semiclassical assumptions \cite{Barrau:2011md}. 
The status of the question has improved with the definition and quantisation of spherical symmetric models
\cite{Gambini:2013nea, Gambini:2013hna, Ghosh:2012wq, Gambini:2013exa}. The approach resembles the hybrid quantisation techniques used in loop quantum cosmology ({\bf Chapter LOOP QUANTUM COSMOLOGY}). More precisely, the quantum spherical background space-time is defined using LQG techniques, whereas perturbations, accounting for Hawking radiation, are described by a quantum test field (defined by means of a Fock Hilbert space).

%

\begin{thebibliography}{104}
\providecommand{\natexlab}[1]{#1}
\providecommand{\url}[1]{\texttt{#1}}
\expandafter\ifx\csname urlstyle\endcsname\relax
  \providecommand{\doi}[1]{doi: #1}\else
  \providecommand{\doi}{doi: \begingroup \urlstyle{rm}\Url}\fi

\bibitem{Narayan:2013gca}
R.~Narayan and J.~E. McClintock, {Observational Evidence for Black Holes}.
  (2013).

\bibitem{Hawking:1971tu}
S.~Hawking, {Gravitational radiation from colliding black holes},
  \emph{Phys.Rev.Lett.} {\bf 26}, \penalty0 1344--1346.

\bibitem{Bekenstein:1973ur}
J.~D. Bekenstein, {Black holes and entropy}, \emph{Phys.Rev.} {\bf D7},
  \penalty0 2333--2346.

\bibitem{Bardeen:1973gs}
J.~M. Bardeen, B.~Carter, and S.~Hawking, {The Four laws of black hole
  mechanics}, \emph{Commun.Math.Phys.} {\bf 31}, \penalty0 161--170.

\bibitem{Hawking:1974sw}
S.~Hawking, {Particle Creation by Black Holes}, \emph{Commun.Math.Phys.} {\bf
  43}, \penalty0 199--220.

\bibitem{Ashtekar:2005qt}
A.~Ashtekar and M.~Bojowald, {Quantum geometry and the Schwarzschild
  singularity}, \emph{Class.Quant.Grav.} {\bf 23}, \penalty0 391--411.

\bibitem{Ashtekar:2005cj}
A.~Ashtekar and M.~Bojowald, {Black hole evaporation: A Paradigm},
  \emph{Class.Quant.Grav.} {\bf 22}, \penalty0 3349--3362.

\bibitem{Ashtekar:2008jd}
A.~Ashtekar, V.~Taveras, and M.~Varadarajan, {Information is Not Lost in the
  Evaporation of 2-dimensional Black Holes}, \emph{Phys.Rev.Lett.} {\bf 100},
  \penalty0 211302.

\bibitem{Ashtekar:2010hx}
A.~Ashtekar, F.~Pretorius, and F.~M. Ramazanoglu, {Surprises in the Evaporation
  of 2-Dimensional Black Holes}, \emph{Phys.Rev.Lett.} {\bf 106}, \penalty0
  161303.

\bibitem{Ashtekar:2010qz}
A.~Ashtekar, F.~Pretorius, and F.~M. Ramazanoglu, {Evaporation of 2-Dimensional
  Black Holes}, \emph{Phys.Rev.} {\bf D83}, \penalty0 044040.

\bibitem{Wald:1999vt}
R.~M. Wald, {The thermodynamics of black holes}, \emph{Living Rev.Rel.} {\bf
  4}, \penalty0 6,  (2001).

\bibitem{Perez:2014ura}
A.~Perez, {Statistical and entanglement entropy for black holes in quantum
  geometry}.  (2014).

\bibitem{Perez:2014xca}
A.~Perez, {No firewalls in quantum gravity: the role of discreteness of quantum
  geometry in resolving the information loss paradox}.  (2014).

\bibitem{Varadarajan:1999it}
M.~Varadarajan, {Fock representations from U(1) holonomy algebras},
  \emph{Phys.Rev.} {\bf D61}, \penalty0 104001.

\bibitem{Ashtekar:2001xp}
A.~Ashtekar and J.~Lewandowski, {Relation between polymer and Fock
  excitations}, \emph{Class.Quant.Grav.} {\bf 18}, \penalty0 L117--L128.

\bibitem{Ashtekar:2002sn}
A.~Ashtekar, S.~Fairhurst, and J.~L. Willis, {Quantum gravity, shadow states,
  and quantum mechanics}, \emph{Class.Quant.Grav.} {\bf 20}, \penalty0
  1031--1062.

\bibitem{ABF}
A.~Ashtekar, C.~Beetle, and S.~Fairhurst, Isolated horizons: a generalization
  of black hole mechanics, \emph{Class. Quant. Grav.} {\bf 16},  (1999).

\bibitem{AFK0}
A.~Ashtekar, C.~Beetle, and S.~Fairhurst, Mechanics of isolated horizons,
  \emph{Class. Quantum Grav.} {\bf 17},  (2000).

\bibitem{ABL1}
A.~Ashtekar, C.~Beetle, and J.~Lewandowski, Mechanics of rotating isolated
  horizons, \emph{Phys. Rev. D}. {\bf 64},  (2001).

\bibitem{AFK}
A.~Ashtekar, S.~Fairhurst, and B.~Krishnan, Isolated horizons: {H}amiltonian
  evolution and the first law, \emph{Phys. Rev. D}. {\bf 62},  (2000).

\bibitem{Lew1}
J.~Lewandowski, Spacetimes admitting isolated horizons, \emph{Class. Quantum
  Grav.} {\bf 17},  (2000).

\bibitem{LP1}
J.~Lewandowski and T.~Paw{\l}owski, Extremal isolated horizons: a local
  uniqueness theorem, \emph{Class. Quantum Grav.} {\bf 20},  (2003).

\bibitem{Kupeli}
D.~N. Kupeli, On null submanifolds in spacetimes, \emph{Geometriae Dedicata}.
  {\bf 23}, \penalty0 33--51,  (1987).

\bibitem{AK2}
A.~Ashtekar and B.~Krishnan, Dynamical horizons and their properties,
  \emph{Phys. Rev. D}. {\bf 68},  (2003).

\bibitem{AKLR}
A.~Ashtekar and B.~Krishnan, Isolated and dynamical horizons and their
  applications, \emph{Living Rev. Relativity}. {\bf 7},  (2004).

\bibitem{Wald}
R.~M. Wald, \emph{General Relativity}. (Chicago University Press, Chicago,
  1984).

\bibitem{Ashtekar:2004gp}
A.~Ashtekar, J.~Engle, T.~Pawlowski, and C.~Van Den~Broeck, {Multipole moments
  of isolated horizons}, \emph{Class.Quant.Grav.} {\bf 21}, \penalty0
  2549--2570.

\bibitem{Ashtekar:2004nd}
A.~Ashtekar, J.~Engle, and C.~Van Den~Broeck, {Quantum horizons and black hole
  entropy: Inclusion of distortion and rotation}, \emph{Class.Quant.Grav.} {\bf
  22}, \penalty0 L27--L34.

\bibitem{GNH}
M.~J. Gotay, J.~N. Nester, and G.~Hinds, Presymplectic manifolds and the
  {D}irac {B}ergmann theory of constraints, \emph{J. Math. Phys.} {\bf 19},
  (1978).

\bibitem{Booth}
I.~S. Booth, Metric-based hamiltonians, null boundaries and isolated horizons,
  \emph{Class. Quantum Grav.} {\bf 18},  (2001).

\bibitem{Witten}
{\v{C}}.~Crnkovi\'c and E.~Witten, Covariant description of canonical formalism
  in geometrical theories (in 300 years of gravitation, {S}. {W}. {H}awking and
  {W}. {I}srael {E}ds.).

\bibitem{Crnk}
{\v{C}}.~Crnkovi\'c, Symplectic geometry of the convariant phase space,
  \emph{Class. Quantum Grav.} {\bf 5},  (1988).

\bibitem{Ashtekar:1999wa}
A.~Ashtekar, A.~Corichi, and K.~Krasnov, {Isolated horizons: The Classical
  phase space}, \emph{Adv.Theor.Math.Phys.} {\bf 3}, \penalty0 419--478,
  (1999).

\bibitem{Engle:2009vc}
J.~Engle, K.~Noui, and A.~Perez, Black hole entropy and {SU}(2)
  {C}hern-{S}imons theory, \emph{Phys. Rev. Lett.} {\bf 105},  (2010).

\bibitem{Engle:2010kt}
J.~Engle, K.~Noui, and A.~Perez, Black hole entropy from the {SU}(2)-invariant
  formulation of type {I} isolated horizons, \emph{Phys. Rev. D}. {\bf 82},
  (2010).

\bibitem{Engle:2011vf}
J.~Engle, K.~Noui, and A.~Perez, The {SU}(2) black hole entropy revisited,
  \emph{JHEP}. {\bf 05},  (2011).

\bibitem{Ashtekar:1998ak}
A.~Ashtekar, A.~Corichi, and J.~A. Zapata, {Quantum theory of geometry III:
  Noncommutativity of Riemannian structures}, \emph{Class.Quant.Grav.} {\bf
  15}, \penalty0 2955--2972.

\bibitem{nos}
J.~F. Barbero~G., J.~Prieto, and E.~J. Villase{\~n}or, {Hamiltonian treatment
  of linear field theories in the presence of boundaries: a geometric
  approach}, \emph{Class.Quant.Grav.} {\bf 31}, \penalty0 045021.

\bibitem{ABK}
A.~Ashtekar, J.~Baez, and K.~Krasnov, {Quantum Geometry of Isolated Horizons
  and Black Hole Entropy}, \emph{Adv.Theor.Math.Phys.} {\bf 4}, \penalty0
  1--94,  (2000).

\bibitem{Domagala:2004jt}
M.~Domaga{\l}a and J.~Lewandowski, Black-hole entropy from quantum geometry,
  \emph{Class. Quant. Grav.} {\bf 21}, \penalty0 5233–--5243,  (2004).

\bibitem{Krasnov:1996tb}
K.~V. Krasnov, {Counting surface states in the loop quantum gravity},
  \emph{Phys.Rev.} {\bf D55}, \penalty0 3505--3513.

\bibitem{Krasnov:1996wc}
K.~V. Krasnov, {On Quantum statistical mechanics of Schwarzschild black hole},
  \emph{Gen.Rel.Grav.} {\bf 30}, \penalty0 53--68.

\bibitem{Krasnov:1997yt}
K.~V. Krasnov, {Quantum geometry and thermal radiation from black holes},
  \emph{Class.Quant.Grav.} {\bf 16}, \penalty0 563--578.

\bibitem{Frodden:2011eb}
E.~Frodden, A.~Ghosh, and A.~Perez, {Quasilocal first law for black hole
  thermodynamics}, \emph{Phys.Rev.} {\bf D87}, \penalty0 121503.

\bibitem{nosLew}
J.~F. Barbero~G., J.~Lewandowski, and E.~J.~S. Villase{\~n}or, Flux-area
  operator and black hole entropy, \emph{Phys. Rev. D}. {\bf 80}, \penalty0
  044016--1--15,  (2009).

\bibitem{Ghosh:2004rq}
A.~Ghosh and P.~Mitra, {A Bound on the log correction to the black hole area
  law}, \emph{Phys.Rev.} {\bf D71}, \penalty0 027502.

\bibitem{Ghosh:2004wq}
A.~Ghosh and P.~Mitra, {An Improved lower bound on black hole entropy in the
  quantum geometry approach}, \emph{Phys.Lett.} {\bf B616}, \penalty0 114--117.

\bibitem{Kaul:1998xv}
R.~K. Kaul and P.~Majumdar, Quantum black hole entropy, \emph{Phys. Lett. B}.
  {\bf 3}, \penalty0 267--270,  (1998).

\bibitem{Ashtekar:1997yu}
A.~Ashtekar, J.~Baez, A.~Corichi, and K.~Krasnov, {Quantum geometry and black
  hole entropy}, \emph{Phys.Rev.Lett.} {\bf 80}, \penalty0 904--907.

\bibitem{Ghosh:2006ph}
A.~Ghosh and P.~Mitra, Counting black hole microscopic states in loop quantum
  gravity, \emph{Phys. Rev. D}. {\bf 74}, \penalty0 064026--1--5,  (2006).

\bibitem{Ghosh:2011fc}
A.~Ghosh and A.~Perez, {Black hole entropy and isolated horizons
  thermodynamics}, \emph{Phys.Rev.Lett.} {\bf 107}, \penalty0 241301.

\bibitem{Ghosh:2013iwa}
A.~Ghosh, K.~Noui, and A.~Perez, {Statistics, holography, and black hole
  entropy in loop quantum gravity}.  (2013).

\bibitem{Corichi:2006bs}
A.~Corichi, J.~Diaz-Polo, and E.~F.-Borja, Quantum geometry and microscopic
  black hole entropy, \emph{Class. Quant. Grav.} {\bf 24}, \penalty0 243--251,
  (2007).

\bibitem{Corichi:2006wn}
A.~Corichi, J.~Diaz-Polo, and E.~F.-Borja, Black hole entropy quantization,
  \emph{Phys. Rev. Lett.} {\bf 98}, \penalty0 181301--1--4,  (2007).

\bibitem{BV1}
J.~F. Barbero~G. and E.~J.~S. Villase{{\~n}}or, Statistical description of the
  black hole degeneracy spectrum, \emph{Phys. Rev. D}. {\bf 83}, \penalty0
  104013--1--21,  (2011).

\bibitem{Bek1}
J.~Bekenstein, Black hole entropy quantization, \emph{Lett. Nuovo Cim.} {\bf
  11}, \penalty0 467--470,  (1974).

\bibitem{Bek2}
J.~Bekenstein and V.~F. Mukhanov, Spectroscopy of the quantum black hole,
  \emph{Phys. Lett. B}. {\bf 360}, \penalty0 7--12,  (1995).

\bibitem{Griffiths}
R.~B. Griffiths, Microcanonical ensemble in quantum statistical mechanics,
  \emph{J. Math. Phys.} {\bf 6}, \penalty0 1447--1461,  (1965).

\bibitem{Meissner:2004ju}
K.~A. Meissner, Black-hole entropy in loop quantum gravity, \emph{Class. Quant.
  Grav.} {\bf 21}, \penalty0 5245–--5251,  (2004).

\bibitem{Agullo:2008yv}
I.~Agull\'o, J.~F. Barbero~G., J.~Diaz-Polo, E.~F.~Borja, and E.~J.~S.
  Villase{\~n}or, Black hole state counting in loop quantum gravity: A
  number-theoretical approach, \emph{Phys. Rev. Lett.} {\bf 100}, \penalty0
  211301--1--4,  (2008).

\bibitem{Agullo:2010zz}
I.~Agull\'o, J.~F. Barbero~G., J.~Diaz-Polo, E.~F.~Borja, and E.~J.~S.
  Villase{\~n}or, Detailed black hole state counting in loop quantum gravity,
  \emph{Phys. Rev. D}. {\bf 82}, \penalty0 084029--1--31,  (2010).

\bibitem{Sahlmann:2007jt}
H.~Sahlmann, Entropy calculation for a toy black hole, \emph{Class. Quant.
  Grav.} {\bf 25}, \penalty0 055004--1--14,  (2008).

\bibitem{Sahlmann:2007zp}
H.~Sahlmann, Toward explaining black hole entropy quantization in loop quantum
  gravity, \emph{Phys. Rev. D}. {\bf 76}, \penalty0 104050--1--7,  (2007).

\bibitem{Agullo:2009zt}
I.~Agullo, E.~F. Borja, and J.~Diaz-Polo, {Computing Black Hole entropy in Loop
  Quantum Gravity from a Conformal Field Theory perspective}, \emph{JCAP}. {\bf
  0907}, \penalty0 016.

\bibitem{BarberoG.:2008ue}
J.~F. Barbero~G. and E.~J.~S. Villase{\~n}or, Generating functions for black
  hole entropy in loop quantum gravity, \emph{Phys. Rev. D}. {\bf 77},
  \penalty0 121502(R)--1--5,  (2008).

\bibitem{G.:2008mj}
J.~F. Barbero~G. and E.~J.~S. Villase{\~n}or, On the computation of black hole
  entropy in loop quantum gravity, \emph{Class. Quant. Grav.} {\bf 26},
  \penalty0 035017--1--22,  (2009).

\bibitem{G.:2011zr}
J.~F. Barbero~G. and E.~J.~S. Villase{\~n}or, The thermodynamic limit and black
  hole entropy in the area ensemble, \emph{Class. Quant. Grav.} {\bf 28},
  \penalty0 215014--1--15,  (2011).

\bibitem{Frodden:2011zz}
E.~Frodden, A.~Ghosh, and A.~Perez, {Black hole entropy in LQG: Recent
  developments}, \emph{AIP Conf.Proc.} {\bf 1458}, \penalty0 100--115.

\bibitem{Wald:1984rg}
R.~M. Wald, {General Relativity}.  (1984).

\bibitem{Thorne:1986iy}
K.~S. Thorne, R.~Price, and D.~Macdonald, {Black holes: the membrane paradigm}.
   (1986).

\bibitem{Carlip:1993sa}
S.~Carlip and C.~Teitelboim, {The Off-shell black hole},
  \emph{Class.Quant.Grav.} {\bf 12}, \penalty0 1699--1704.

\bibitem{Ghosh:2008jc}
A.~Ghosh and P.~Mitra, {Fine-grained state counting for black holes in loop
  quantum gravity}, \emph{Phys.Rev.Lett.} {\bf 102}, \penalty0 141302.

\bibitem{FernandoBarbero:2011kb}
G.~Fernando~Barbero and E.~J. Villase{\~n}or, {Statistical description of the
  black hole degeneracy spectrum}, \emph{Phys.Rev.} {\bf D83}, \penalty0
  104013.

\bibitem{Jacobson:1995ab}
T.~Jacobson, {Thermodynamics of space-time: The Einstein equation of state},
  \emph{Phys.Rev.Lett.} {\bf 75}, \penalty0 1260--1263.

\bibitem{Smolin:2012ys}
L.~Smolin, {General relativity as the equation of state of spin foam}.  (2012).

\bibitem{Jacobson:2007uj}
T.~Jacobson, {Renormalization and black hole entropy in Loop Quantum Gravity},
  \emph{Class.Quant.Grav.} {\bf 24}, \penalty0 4875--4879.

\bibitem{'tHooft:1984re}
G.~'t~Hooft, {On the Quantum Structure of a Black Hole}, \emph{Nucl.Phys.} {\bf
  B256}, \penalty0 727.

\bibitem{Frodden:2012dq}
E.~Frodden, M.~Geiller, K.~Noui, and A.~Perez, {Black Hole Entropy from complex
  Ashtekar variables}.  (2012).

\bibitem{Han:2014xna}
M.~Han, {Black Hole Entropy in Loop Quantum Gravity, Analytic Continuation, and
  Dual Holography}.  (2014).

\bibitem{Achour:2014eqa}
J.~B. Achour, A.~Mouchet, and K.~Noui, {Analytic Continuation of Black Hole
  Entropy in Loop Quantum Gravity}.  (2014).

\bibitem{Geiller:2013pya}
M.~Geiller and K.~Noui, {BTZ Black Hole Entropy and the Turaev-Viro model}.
  (2013).
\newblock \doi{10.1007/s00023-014-0331-7}.

\bibitem{Ghosh:2014rra}
A.~Ghosh and D.~Pranzetti, {CFT/Gravity Correspondence on the Isolated
  Horizon}.  (2014).

\bibitem{Pranzetti:2013lma}
D.~Pranzetti, {Black hole entropy from KMS-states of quantum isolated
  horizons}, \emph{Phys.Rev.} {\bf D89}, \penalty0 104046.

\bibitem{Frodden:2012nu}
E.~Frodden, M.~Geiller, K.~Noui, and A.~Perez, {Statistical Entropy of a BTZ
  Black Hole from Loop Quantum Gravity}, \emph{JHEP}. {\bf 1305}, \penalty0
  139.

\bibitem{Hawking:1976de}
S.~Hawking, {Black Holes and Thermodynamics}, \emph{Phys.Rev.} {\bf D13},
  \penalty0 191--197.

\bibitem{Bianchi:2012vp}
E.~Bianchi and W.~Wieland, {Horizon energy as the boost boundary term in
  general relativity and loop gravity}.  (2012).

\bibitem{Unruh:1976db}
W.~Unruh, {Notes on black hole evaporation}, \emph{Phys.Rev.} {\bf D14},
  \penalty0 870.

\bibitem{Huszar:1971nn}
M.~Huszar, {Angular momentum and unitary spinor bases of the lorentz group},
  \emph{Acta Phys.Hung.} {\bf 30}, \penalty0 241--251.

\bibitem{Chirco:2014saa}
G.~Chirco, H.~M. Haggard, A.~Riello, and C.~Rovelli, {Spacetime thermodynamics
  without hidden degrees of freedom}.  (2014).

\bibitem{Solodukhin:2011gn}
S.~N. Solodukhin, {Entanglement entropy of black holes}, \emph{Living Rev.Rel.}
  {\bf 14}, \penalty0 8,  (2011).

\bibitem{Bianchi:2012br}
E.~Bianchi, {Horizon entanglement entropy and universality of the graviton
  coupling}.  (2012).

\bibitem{Bianchi:2013rya}
E.~Bianchi and A.~Satz, {Mechanical laws of the Rindler horizon},
  \emph{Phys.Rev.} {\bf D87}, \penalty0 124031.

\bibitem{Wald:1995yp}
R.~M. Wald, {Quantum field theory in curved space-time and black hole
  thermodynamics}.  (1995).

\bibitem{Casini:2008cr}
H.~Casini, {Relative entropy and the Bekenstein bound},
  \emph{Class.Quant.Grav.} {\bf 25}, \penalty0 205021.

\bibitem{Bianchi:2012ui}
E.~Bianchi, {Entropy of Non-Extremal Black Holes from Loop Gravity}.  (2012).

\bibitem{Bianchi:2012ev}
E.~Bianchi and R.~C. Myers, {On the Architecture of Spacetime Geometry}.
  (2012).

\bibitem{Gibbons:1976ue}
G.~Gibbons and S.~Hawking, {Action Integrals and Partition Functions in Quantum
  Gravity}, \emph{Phys.Rev.} {\bf D15}, \penalty0 2752--2756.

\bibitem{York:1986it}
J.~York, James~W., {Black hole thermodynamics and the Euclidean Einstein
  action}, \emph{Phys.Rev.} {\bf D33}, \penalty0 2092--2099.

\bibitem{Sen}
A.~Sen, {Logarithmic Corrections to Schwarzschild and Other Non-extremal Black
  Hole Entropy in Different Dimensions}, \emph{JHEP}. {\bf 1304}, \penalty0
  156.

\bibitem{Barrau:2011md}
A.~Barrau, T.~Cailleteau, X.~Cao, J.~Diaz-Polo, and J.~Grain, {Probing Loop
  Quantum Gravity with Evaporating Black Holes}, \emph{Phys.Rev.Lett.} {\bf
  107}, \penalty0 251301.

\bibitem{Gambini:2013nea}
R.~Gambini and J.~Pullin, {Hawking radiation from a spherical loop quantum
  gravity black hole}.  (2013).

\bibitem{Gambini:2013hna}
R.~Gambini, J.~Olmedo, and J.~Pullin, {Quantum black holes in Loop Quantum
  Gravity}, \emph{Class.Quant.Grav.} {\bf 31}, \penalty0 095009.

\bibitem{Ghosh:2012wq}
A.~Ghosh and A.~Perez, {The scaling of black hole entropy in loop quantum
  gravity}.  (2012).

\bibitem{Gambini:2013exa}
R.~Gambini and J.~Pullin, {An introduction to spherically symmetric loop
  quantum gravity black holes}.  (2013).

\end{thebibliography}

\printindex                         
\end{document}